\documentclass[vecphys]{svmult}  
\usepackage[dvips]{graphicx}
\usepackage{amssymb}
\usepackage{makeidx}
\usepackage{multicol}
\usepackage[bottom]{footmisc}
\usepackage{cite}
\makeindex
\newcommand{\lapproxeq}{\lower .7ex\hbox{$\;\stackrel{\textstyle  
<}{\sim}\;$}} 
\newcommand{\gapproxeq}{\lower .7ex\hbox{$\;\stackrel{\textstyle  
>}{\sim}\;$}} 
\newcommand{\stackdown}[2]{\lower 1.4ex\hbox{$\;\stackrel{\textstyle{#1}}  
{\scriptstyle{#2}}\;$}}
\newcommand{\be}{\begin{equation}} 
\newcommand{\ee}{\end{equation}} 
\newcommand{\bea}{\begin{eqnarray}} 
\newcommand{\eea}{\end{eqnarray}}

\def\laq{~\raise 0.4ex\hbox{$<$}\kern -0.8em\lower 0.62
ex\hbox{$\sim$}~}

\def\gaq{~\raise 0.4ex\hbox{$>$}\kern -0.7em\lower 0.62
ex\hbox{$\sim$}~}

\def \a {\alpha}

\def \Ga {\Gamma}

\def \sg {\sigma}

\def \r {\rho}

\def \wt {\widetilde}

\newcommand{\epsi}{e^\psi}
\newcommand{\rhm}{\rho_m}

\makeatletter 
\def\slash{\@ifnextchar[{\fmsl@sh}{\fmsl@sh[0mu]}} 
\def\fmsl@sh[#1]#2{%
  \mathchoice 
    {\@fmsl@sh\displaystyle{#1}{#2}}%
    {\@fmsl@sh\textstyle{#1}{#2}}%
    {\@fmsl@sh\scriptstyle{#1}{#2}}%
    {\@fmsl@sh\scriptscriptstyle{#1}{#2}}} 
\def\@fmsl@sh#1#2#3{\m@th\ooalign{$\hfil#1\mkern#2/\hfil$\crcr$#1#3$}} 
\makeatother 
\begin{document}
\title*{The issue of Dark Energy in String Theory}
\author{Nick~E.~Mavromatos}
\institute{King's College London, Department of Physics, Theoretical 
Physics,   
Strand , London WC2R 2LS, U.K.\\
\texttt{Nikolaos.Mavromatos@kcl.ac.uk}}
\maketitle
Recent astrophysical observations, pertaining to 
either high-redshift supernovae or 
cosmic microwave background temperature fluctuations, as those 
measured recently by the WMAP satellite, 
provide us with data of unprecedented accuracy, 
pointing towards two (related) facts: 
(i) our Universe is accelerated at present, and (ii) more than 70 \% of its 
energy content consists of 
an unknown substance, termed dark energy, which is believed responsible
for its current acceleration. Both of these facts are a challenge 
to String theory. In this review I outline briefly the 
challenges, the problems and possible avenues for research towards a 
resolution of the Dark Energy issue in string theory.   

\section{Introduction}

Recent Astrophysical Data, from either studies of distant 
supernovae type Ia~\cite{snIa}, or precision measurements of temperature 
fluctuations in the cosmic microwave background radiation from the 
WMAP satellite~\cite{wmap}, point towards a current-era acceleration of 
our Universe, as well as a very peculiar energy budget for it,
70\% of the energy density of which consists of an unknown energy substance,
termed dark Energy.  
In fact, global best-fit models of a compilation of all the available data
at present are provided by simple Einstein-Friedman Universes with a 
(four space-time dimensional) positive {\it cosmological constant} 
$\Lambda$, whose value saturate the 
Newtonian upper limit obtained from galactic dynamics, 
namely in order of magnitude 
\begin{equation}
\Lambda \sim 10^{-122}M_P~\qquad (M_P = 10^{19}~{\rm GeV}~).   
\label{lambdaval} 
\end{equation} 
Although, as a classical (general relativistic) field theory, 
such a model is fairly simple, from a quantum theory view point 
it appears to be the less understood at present. The reason is simple: 
Since in cosmology~\cite{cosmology} the radiation and matter energy densities
scale with inverse powers of the scale factor, $a^{-4}$ and $a^{-3}$ 
respectively, in a Universe with positive cosmological constant $\Lambda$, 
the vacuum energy density 
remains constant and positive, and eventually dominates the energy budget. 
The asymptotic (in time) Universe becomes a {\it de Sitter} one, 
and in such a Universe the scale factor will increase exponentially, 
\begin{equation} 
a(t) = a_0 e^{\sqrt{\frac{\Lambda}{3}}t}~,
\label{inflation} 
\end{equation}
thereby implying that the Universe will eventually enter an inflationary 
phase again, and in fact it will accelerate eternally, since 
${\ddot a} > 0$, where  
the overdot denotes derivative with respect to 
the Robertson-Walker cosmic time, $t$, defined by: 
\begin{equation} 
ds_{RW}^2 = -dt^2 + a^2(t)ds_{\rm spatial}^2 
\label{rw}
\end{equation}
In such de Sitter Universes there is 
unfortunately a {\it cosmic horizon} 
\begin{equation}
\delta \propto \int_{t_0}^{t_{End}} \frac{cdt}{a(t)} < \infty 
\label{horizon}
\end{equation}
where $t_{End}$ 
indicates the end of time. For a closed Universe $t_{End} < \infty$,
but for an open or flat Universe $t_{End} \to \infty$.
The Cosmic Microwave (CMB) data of WMAP and other experiments at present 
indicate that our Universe is 
spatially {\it flat}, and hence $t_{End} \to \infty$.

The presence of a cosmic horizon implies that it is not possible to define
pure state vectors of quantum  asymptotic (in time) states. Therefore, 
the entire concept of a well-defined and gauge invariant Scattering matrix
$S$ breaks down in quantum field theories defined on such de Sitter 
space time backgrounds. For string theory 
this is bad news, because precisely 
by construction~\cite{strings}, perturbative string theory is based 
on the well-defined nature of scattering amplitudes of various excitations, 
and hence on a well-defined S-matrix~\cite{smatrix}.  
This is a challenge for string theory, and certainly one of the most important 
issues I would like to discuss in this brief review. 

A straightforward way out, would be 
{\it quintessence}-like scenaria for dark energy~\cite{quint}, 
according to which the latter is due to a potential of a time dependent
scalar field,
which has not yet reached its equilibrium point. If, then,  
the asymptotic value of the 
dark energy vanishes in such a way so 
as not to have a cosmic horizon, then the model could 
be accommodated within string-inspired effective field theories, 
and could thus characterise the 
low-energy limit of strings, given that 
an asymptotic S-matrix could be defined in such a case. 

However, this does not mean that de Sitter Universes {\it per se} cannot 
be accommodated somehow into a (possibly non perturbative) string theory 
framework. Their anti-de-Sitter (AdS, 
negative cosmological constant) counterparts
certainly do, and in fact there have been important development towards
a holographic property of quantum field theories in such Universes, 
due to the celebrated Maldacena conjecture~\cite{malda}, concerning 
quantum properties of (supersymmetric) conformal field theories on 
the boundary of AdS space time.
As we shall discuss  in the next section, similar 
conjectures~\cite{strom} 
may characterise their de Sitter counterparts, 
and this may be a way forward to accommodate such a space time into 
string theory. 

Finally, a more straightforward (perturbative) 
approach to discuss de Sitter and 
inflationary scenaria in string theories, will be to use the so-called 
non-critical (or Liouville) string framework~\cite{ddk}, 
dealing with a mathematically consistent way of discussing 
strings propagating in non-conformal backgrounds, such as the de Sitter 
space time. This theory, however, at least as far as computation of
the pertinent correlation functions are concerned, 
has not been developed to the same level of mathematical understanding as the 
critical strings. A crucial ingredient in this approach is 
the identification 
of the Liouville mode with the target time~\cite{emn}, which allows for some 
non-conformal backgrounds in string theory, including de Sitter 
space times and accelerated Universes, to be accommodated
in a mathematically consistent manner.

We should stress at this point, that 
the above considerations, regarding S-matrix amplitudes in de Sitter 
Universes, refer to pure perturbative string theories. In the modern approach 
to string theory, where membrane (D-brane) structures~\cite{membranes}
also appear as mathematically consistent entities, the presence of 
a dark energy on the string theory on the brane is unavoidable,
unless extreme conditions on unbroken supersymmetry and static nature of 
brane worlds are imposed. However, in brane cosmology one needs moving branes,
in order to obtain a cosmological space time~\cite{binetruy}, 
and in this case, target space time supersymmetry breaks down, due to 
the brane motion, resulting in non-trivial 
vacuum energy contributions on the brane~\cite{maartens}.

The structure of the article will be the following: 
in section 2, I will deal with mathematical 
properties of de Sitter space times: after  
reviewing briefly basic features of this geometry, 
I will describe modern approaches to the issue of placing a quantum field 
theory in de Sitter space times, 
by discussing briefly a holographic conjecture, 
put forward by Strominger~\cite{strom}, 
according to which a quantum field theory 
on the single boundary of de Sitter space can be related to 
a classical theory in the bulk, 
in a way not dissimilar to the celebrated Maldacena conjecture~\cite{malda} 
for anti-de Sitter spaces (negative cosmological constant space times).
In section 3, I will discuss the issue of cosmic horizons 
in perturbative string theory, and give 
further arguments that consistent perturbative strings 
cannot be characterized by such horizons. 
In section 4, I will discuss quintessence scenaria in strings, where the 
dilaton behaves as the quintessence field, 
responsible for the current acceleration 
of the Universe. I will discuss two opposite examples, 
a pre Big-Bang scenario~\cite{veneziano}, 
in which the string coupling increases at late times, 
with string loop corrections playing a dominant r\^ole, and 
another scenario~\cite{emn,dgmpp}, in which the 
string coupling becomes more and more perturbative as the time 
passes, leading asymptotically to a vanishing dark 
energy, in such a way that S-matrix states can be defined.
In this second scenario the current-era acceleration parameter turns out to be 
proportional to the square of the string coupling, which at present enjoys 
perturbative values compatible with particle physics phenomenology.   
I will briefly discuss predictions 
of such models in the context of recent data, but also unresolved problems. 
I will not discuss the issue of dark energy in brane 
cosmologies in this article, 
as this is a topic covered by other lecturers in the school~\cite{maartens}.
Conclusions and directions for future research in the issue of Dark 
Energy in Strings will be presented in section 5. 

\section{De Sitter (dS) Universes from a Modern Perspective}

\begin{figure}[t]
\centering
\includegraphics[width=5.7cm]{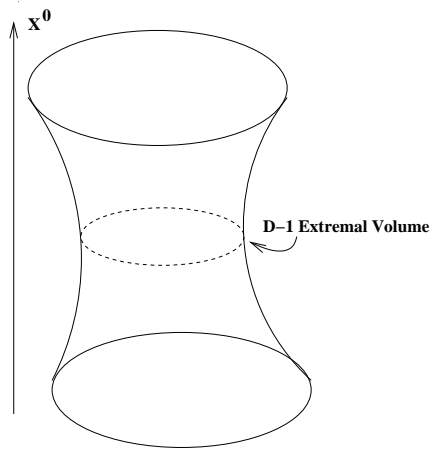} 
\caption { A de Sitter space is a single-sheet hyperboloid,
obtained from a (d+1)-dimensional Minkowski space-time 
by an appropriate embedding of the hypersurface 
$ - (X^0)^2 + (X^1)^2 + \dots + (X^d)^2 = \ell ^2 $, where 
$\ell$ is the de Sitter radius.} 
\label{fig:ds}
\end{figure}

In this section we shall give a very brief overview of the 
most important properties of de Sitter space, relevant for 
our discussion. For more details we refer the reader to 
\cite{strom}, and references therein, 
where a concise exposition of the most important  
properties of classical and quantum theories of de Sitter space
is given.

\subsection{Classical Properties} 

The classical Geometrical picture of a de Sitter space time 
is that of a single-sheet hyperboloid, depicted in figure~\ref{fig:ds}.
This hypersurface can be constructed from the flat (d+1)-dimensional
Minkowski space time, with coordinates $(X^0,X^i)~, i=1, \dots d$,  
by means of the equation:
\be 
 - (X^0)^2 + (X^1)^2 + \dots + (X^d)^2 = \ell ^2 
\label{dsdef}
\ee
where the parameter $\ell$ has units of length, and is called the 
{\it de Sitter radius}.

The classical  Einstein equations, which yield as a solution 
this space time, involve a {\it positive cosmological constant }
\be 
   R_{\mu\nu} - \frac{1}{2}g_{\mu\nu} + \Lambda g_{\mu\nu} = 0~, \qquad 
\Lambda = \frac{(d-2)(d-1)}{2\ell^2}
\label{einsteq}
\ee
There are various coordinates one can use for the description 
of such space times, whose detailed description can be found in \cite{strom}.
The most useful one, and most relevant for our purposes, 
which helps us understand the causal properties of 
the de Sitter space time, are the conformal coordinates, 
$(T, \theta_i)~, i=1, \dots d$, in terms of which the metric element reads:
\be
ds^2 = \frac{1}{{\rm cos}^2T}\left(-dT^2 + d\Omega_{d-1}^2\right)
\label{conf}
\ee 
where $\Omega$ is the usual angular part, expressed in terms of $\theta_i$'s.

\begin{figure}[t]
\centering
\includegraphics[width=12.0cm]{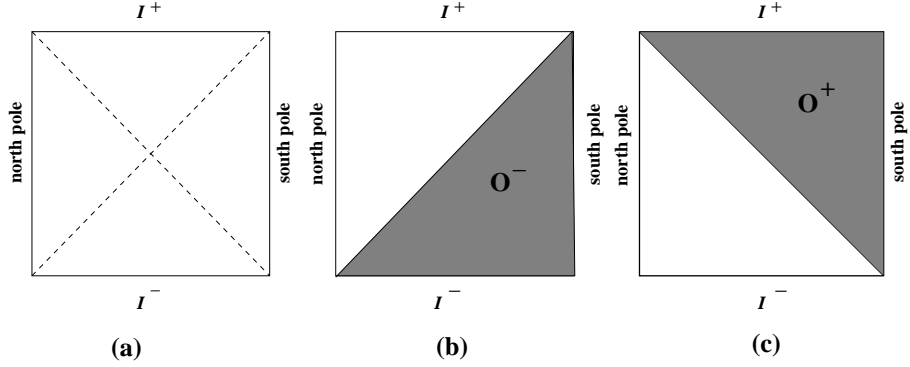} 
\caption { (a): The Penrose diagram for the de Sitter space, constructed
by means of conformal coordinates. Horizontal slices represent 
the extremal volume 
$S^{d-1}$ (c.f. dashed line in fig.~\ref{fig:ds}), 
whilst every point in the interior represents a $S^{d-2}$. 
The vertical slices marked as north and south pole are time-like surfaces.
The $I^+$ ($I^-$) surfaces correspond to the future (past) infinity, and they are the surfaces where all the null geodesics ($ds^2 =0$) originate and 
terminate. The diagonal doted lines represent the past and future horizons
of an observer at the south pole.
Due to the existence of an horizon in this geometry,
a light that starts at, say, the north pole at $I^-$ will reach the south pole 
by the time it reaches $I^+$ {\it infinitely far} in the future. 
(b) A classical observer sitting at the south pole will never be able 
to observe anything past the dotted line that stretches from  
the north pole at $I^-$ to the south pole at $I^+$ (causal past 
region ${\cal O}^-$). 
(c) Similarly, the south pole observer will only be able to send a 
message {\it only} to the causal future region ${\cal O}^+$.}
\label{fig:penrose}
\end{figure}

In terms of these coordinates, one arrives easily at the Penrose diagram
for the de Sitter space, depicted in fig.~\ref{fig:penrose}(a), 
which contains
all the information about the causal structure.

A peculiar feature of this space, but quite important for 
the development of a consistent 
quantum gravity theory, is the fact that 
{\it no single observer can access the entire space time} 
(see figs.~\ref{fig:penrose}(b),(c)). 
As we see from the figure, the causal past and future regions 
of an observer sitting, say, at the south pole will only be 
the portions ${\cal O}^-$ and ${\cal O}^+$, respectively.
Their intersection (called {\it causal diamond}) is the only {\it 
region} of the de Sitter space that is fully accessible 
to an observer at the south pole.

\subsection{Quantum Field Theory on dS: thermodynamical properties} 

Due to the aforementioned problems of inaccessibility of the entire 
region of dS space to a classical observer, a consistent 
formulation of quantum field theory in such space times is still
an open issue, and certainly it is expected to be rather different 
from the corresponding one in Minkowskian space times, where such 
inaccessibility problems are absent.

The situation somewhat resembles that of a Black Hole (BH).
In that problem, there is an horizon, for an asymptotic observer,
who lies far away from it and makes his/her measurements
locally. The entropy $S$ associated with an event horizon in the 
BH case is given by the Bekenstein-Hawking area law~\cite{entropy,hawking}
\be
S = \frac{1}{4G_N}A
\label{entropy}
\ee
where $G_N$ is the gravitational (Newton) constant, and $A$ is the area of the 
horizon. This is a macroscopic formula, which essentially 
describes how properties of event horizons in General Relativity 
change as their parameters are varied. 

The quantum BH Hawking radiate, and from this type of particle creation
one comes to the conclusion that there is a temperature $T_H$ characterising 
the exterior space time (`Hawking' temperature), measured at infinity.
For the case of a Schwarzschild BH
\be 
T_H = \frac{\hbar}{8\pi M} 
\label{th}
\ee
where $M$ is the mass (energy) of the BH. In the BH case this temperature
is found by integrating the thermodynamic relation $dS_{BH}/dM = 1/T_{H}$.

From a quantum field theory point of view, we can understand such formulae 
from the fact that they describe some effective ``loss'' of information, 
associated with modes that go beyond the horizon, and hence are lost for ever
for the classical asymptotic observer. 

Indeed, if one considers the exterior portion of space time of the 
BH as an open system, and the interior as constituting the ``environment'',
with which the physical world interacts, 
then the Bekenstein-area law formula may be derived simply even in 
flat Minkowski space times with a boundary of area $A$.
For instance, Srednicki~\cite{srednicki} has demonstrated that
by tracing the density matrix of a massless scalar field (taken as a
toy, but illustrative, example) over degrees of freedom 
residing inside an imaginary sphere, embedded in a flat Minkowski space time,
the result leads to an entropy for the scalar field which is 
proportional to the area, and {\it not} the volume of the sphere.

In view of this analogy, one would therefore expect that in all cases,
where there is a region in space time inaccessible to an asymptotic 
observer of a quantum field theory on such geometries, 
there should be an entropy associated with the area of the region.
This should also be expected in the dS case, in view of the existence 
of a de Sitter horizon. If one postulates some thermodynamic 
properties associated with the entropy, one arrives also at an effective 
temperature concept. 

The deep issue in the black hole case is to understand the precise 
coefficient $1/4G_N$ in Eq. (\ref{entropy}), in other words 
develop a sufficiently correct quantum theory for such space times,
in which it will be possible to count the {\it microstates} of the BH 
exactly.  If the latter are associated to a {\it Von Neumann} entropy,
$S_{VN} = -k_B{\rm Tr}\rho {\rm ln \rho} $, where $\rho $ is 
the density matrix
of the system under consideration (the quantum BH in our case),  
and the ${\rm Tr}$ is taken over all microstates, then one should show that
$S_{VN} = S $ given by (\ref{entropy}). 

At present this is one of the most important issues in theoretical 
physics. A precise counting of microstates, however, 
leading to a relation of the form (\ref{entropy}) has become 
possible for certain highly supersymmetric black hole backgrounds 
in string theory, saturating the so-called Bogomolnyi-Prasad-Sommerfeld 
(BPS) bound~\cite{bps}. It is, though, still unproven 
for general, non supersymmetric black holes, which are the likely 
types to be encountered in our physical world. 

We now come to the dS case. Indeed, as one would expect from 
the above generic arguments, there should be an entropy associated with the 
horizon. In fact it is, and 
there is also a temperature (`Gibbons-Hawking temperature')~\cite{gh}, 
in complete
analogy with the BH case. 
In fact, the Temperature is given in terms of the de Sitter 
radius by:
\be 
  T_{GH} = \frac{1}{2\pi \ell} 
\label{ghtemp}
\ee
and the entropy, associated with the de Sitter horizon of area $A$, 
is given exactly by the formula (\ref{entropy}). 

These properties can 
be proven by considering a quantum field on the dS background and 
evaluating its Green functions.
Such an analysis shows that, in the case of massive quantum fields, 
an observer, moving along a time-like geodesic of dS space, 
observes a thermal bath of particles, when the massive field is in 
its vacuum state $|0>$. It turns out that the correct type of Green 
functions to be used in this case are the thermal ones. 
For details we refer the reader to 
the lectures by Strominger~\cite{strom}, and references therein.
Such an analysis allows for the computation of the effective 
temperature the dS space is associated with.

The entropy of the de Sitter space 
$S_{dS}$, then, is found following the argument suggested by 
Gibbons and Hawking~\cite{gh}, according to which 
\be 
\frac{dS_{dS} }{d(-E_{dS})} = \frac{1}{T_{GH}}
\label{dsentropy}
\ee
where $E_{dS}$ is the energy of the dS space. 
Notice the minus sign in front of $E_{dS}$. This stems from the fact 
that what we call energy in dS space is not as simple as the mass of the 
BH case. To understand qualitatively what might happen in the dS case,
we should first start from the principle outlined above, that the entropy 
of the space is associated with ``stuff'' behind the horizon. We do not, 
at present, have any idea what the ``microstates'' of the dS vacuum are, 
but let us suppose for the sake of the argument, that an entropy is 
associated with them (this assumption is probably correct). 

In general relativity energy is defined as an integral of a 
total derivative over a space-time volume, which therefore reduces 
to a surface integral on the boundary of the surface, 
and hence vanishes for a closed surface. Because of this 
vanishing result, if we consider a closed surface on de Sitter space, 
and we put, say, positive energy on the south pole, then there 
must be necessarily some negative energy 
at the north pole to compensate, and yield a zero result. 

Therefore, the singularity at the north pole, behind the dS horizon,
will correspond to negative energy. 
From the BH analogy, it is therefore more sensible to vary with respect 
to this negative energy, and this explains the relative minus sign
in (\ref{dsentropy})
yielding the correct expression for the area law in the dS case.  

The important point to notice, however, is that, despite the formal similarity 
of the dS with the BH, in the former case no one understands, at present, the precise {\it microscopic} origin of the entropy and temperature. 
It is not clear what precisely are the microstates behind the horizon,
which constitute the ``environment'' with which the quantum field theory 
interacts. 

This question acquires much bigger importance in cosmologies with 
a positive cosmological constant,  
which are currently favoured 
by the astrophysical data~\cite{snIa,wmap}.
Indeed in such cases, the asymptotic (in time) Universe will enter a 
pure de-Sitter-space phase, since all the matter energy density 
will be diluted, 
scaling with the scale 
factor as $a^{-3}$, thereby leaving us only with the 
constant vacuum energy contribution $\Lambda$.  
As discussed in the beginning of the lecture, 
the cosmological horizon will
be given by (\ref{horizon}), and in this case the 
dS radius $\ell$, in terms of which the entropy and temperature are expressed, 
is associated with $\Lambda$ by (\ref{einsteq}), essentially its square root.

\subsection{Lack of Scattering Matrix and intrinsic CPT Violation in dS?}

The important question, therefore, from a quantum-field-theory viewpoint 
on such cosmologies and in general dS-like space times, 
concerns the kind of quantum field theories one can define consistently 
in such a situation.
In this respect, the situation is dual to the BH case in the following 
sense: in a BH, there is an horizon which defines a space time boundary for 
an asymptotic observer who lies far away from it. In a full 
quantum theory the BH evaporates due to Hawking radiation.
Although the above thermodynamics arguments are valid for 
large semi-classical BH, one expects the Hawking evaporation 
process to continue until the BH acquires a size comparable to the characteristic scale of quantum gravity (QG), the Planck length $\ell_P=1/M_P$, with $M_P \sim 10^{19}$ GeV. Such {\it microscopic } BH may either evaporate completely, 
leaving behind a naked singularity, or, better -thus satisfying 
the cosmic censorship hypothesis, according 
to which there are no unshielded space-time singularities
in the physical world - disappear in a space-time ``foam'',
namely in a QG ground state, consisting of dynamical ``flashing on and off''
microscopic BH. In such a case, an initially pure 
quantum state will in principle 
be observed as mixed by the asymptotic observer, 
given that ``part of the state quantum numbers'' will be kept inside
the foamy black holes (``effective information loss''), and hence these
will constitute degrees of freedom inaccessible to the observer.

Barring the importing concept of {\it holographic} properties, 
which may indeed characterise such singular space times in QG, 
to which we shall come later on, 
a situation like this will imply an effective non-unitary evolution 
of quantum states of matter in such backgrounds, and hence
gravitational decoherence. 

A similar situation will characterise the dS space, which 
is dual to the BH analogue, in the sense that the 
observer is inside the (cosmological) horizon, in contrast to the 
BH where he/she was lying outside. However the situation concerning
the inability to define asymptotically pure state vectors for the 
quantum state of matter fields remains in this case. 

The lack of a proper definition of pure ``out'' state asymptotic vectors
in {\it both} situations, 
implies that a gauge invariant scattering matrix is also ill defined in
the dS case.
By a theorem due to Wald then~\cite{wald}, one cannot define 
in such 
quantum field theories a quantum mechanical CPT operator. This 
leads to quantum decoherence of matter propagating in such de Sitter
space times.
For more details I refer the interested reader in \cite{mavrodecoh},
where possible phenomenological consequences of such decoherence are 
discussed in detail.  

\subsection{Holographic properties of dS? Towards a quantum gravity theory} 

A final, but important aspect, that might characterise 
a quantum theory in de Sitter 
space times, is the aforementioned property of holography.
If this happens, then the above-mentioned 
information loss paradox will not occur, and a mathematically consistent 
quantum mechanical picture of gravity in the presence of 
space-time boundaries  
will be in place. 

I must stress, at this point, an 
important issue for which there is often confusion in the 
literature. If quantum gravity turns out to lead to 
open-system quantum mechanics for matter theories, this is 
not necessarily a mathematical inconsistency. It simply means that 
there is information carried out by the quantum-gravitational 
degrees of freedom, which however may not be easy to retrieve in 
a perturbative treatment. Of course, even in such situations, the 
complete system, gravity plus matter, is mathematically a closed quantum 
system. On the other hand, if holography is 
valid, then one simply does not have to worry about any effective 
loss of information due to the space time boundary, and 
hence the situation becomes much cleaner.

Holographic properties of anti-de-Sitter (AdS) spaces (negative 
cosmological constant) are encoded in the 
celebrated Maldacena conjecture~\cite{malda}, according to which 
the quantum correlators of a conformal quantum field theory on the boundary of 
the AdS space can be evaluated by means of classical gravity in the bulk
of this space. This conjecture, known 
with the abbreviation AdS/CFT correspondence, 
has been verified to a number of highly 
supersymmetric backgrounds in string theory, but of course it may not
be valid in (realistic) 
non conformal, non supersymmetric cases. The issue for such 
cases is still open.

A similar conjecture in de Sitter space times has been put forward by 
Strominger~\cite{strom}. The conjecture, which is not proven at present, 
can be formulated as follows: 

Consider an operator $\phi (x_i)$ of quantum gravity in a de Sitter 
space, inserted at points $x_i $ on the hypersurfaces $I^-$ or $I^+$.
The dS/CFT  conjecture  states that correlation functions 
of this operator at the points $x_i$ can be generated by an 
appropriate Euclidean conformal field theory 
\be
\langle \phi(x_1) \dots \phi(x_i) \rangle_{{\rm d}S^{d+1}} \leftarrow\rightarrow
\langle {\cal O}_\phi (x_1) \dots {|cal O}_\phi (x_i) 
\rangle_{S^{d}}
\label{dscft}
\ee
where ${\cal O}_\phi (x)$ is an operator of the CFT associated 
with the operator $\phi$. 

For the simple, but quite instructive case, of a three dimensional 
dS$_3$ space, a proof of this correspondence has been given 
in \cite{strom}, making appropriate use of properties of the 
asymptotic symmetry group of gravity for dS$_3$. 
We refer the interested reader to that work, and references
therein, for more details.

Before closing this section, we would like to stress that 
the dS/CFT conjecture may not be valid in realistic cosmologies,
in which the quantum field theories of relevance are certainly not conformal.
If, however, this conjecture is valid, then this is 
a very big step towards a CPT invariant, non-perturbative, construction 
of a quantum theory of gravity.

The holographic principle~\cite{holsussk} will basically allow for any 
possible information loss associated with the 
presence of the cosmological horizon 
to decay with the cosmic time, in such a way that
an asymptotic observer will not eventually loose any information.
This will allow for a consistent CPT operator to be defined, then,
according to the above-mentioned theorem of Wald~\cite{wald}.

If true for the dS case, one expects a similar holographic property 
to be valid for the BH case as well. 
In fact recently, Hawking argued~\cite{hawk2} this to be 
the case in a BH quantum theory of gravity,
but in my opinion his arguments are not supported by any 
rigorous calculation. Hawking's argument is based on the 
fact that any consistent theory of gravity should involve an appropriate
sum over topologies, including the Minkowskian one (trivial). 
In Hawking's argument, then, the Euclidean path integrals 
over the non-trivial topologies, 
that would give non-unitary contributions, and hence  
information loss, lead to 
expressions in scattering amplitudes 
that decay exponentially with time, thereby leaving only
the trivial topology contributions, which are unitary.   
As we said, however, there is no rigorous computation involved
to support this argument, at least at present,
not withstanding the fact that the Euclidean formalism seems crucial
to the result (although, arguably we know of no other way of performing 
a proper quantum gravity path integral).
Hence, the issue of unitarity 
in effective low-energy theories of quantum gravity 
is still wide open in my opinion, and constitutes a
challenge for both theory and phenomenology of quantum 
gravity~\cite{mavrodecoh}.

\section{No Horizons in Perturbative (Critical) String Theory} 

As discussed above, 
if holography is valid, 
there should, in principle, be no issue regarding string theory and
CPT would be a good symmetry of the theory, as seems desirable
from a modern M-theory point of view~\cite{dineM}.

If, however, holography is not valid for realistic 
non-supersymmetric, non-conformal theories, 
then such a situation is most problematic in string 
theory, which, as mentioned in the beginning, at least in its 
perturbative treatment is based on a formalism with 
{\it well-defined} scattering amplitudes~\cite{smatrix}. 

Apart from the 
scattering-matrix and CPT-based issues,
there are other arguments 
 that exclude the existence of horizons in perturbative string 
theory~\cite{banks}. These arguments derive from considerations
of the shape of the potentials 
arising from supersymmetry breaking scenarios
in perturbative string theory, whose coupling (before compactification)
is defined by the exponential of the dilaton field $g_s= e^\Phi$.

The situation becomes cleanest if we consider, for simplicity and 
definiteness, the case
of a single scalar, canonically normalised, field $\phi$, 
playing the r\^ole of the quintessence field in a 
Robertson-Walker space time with scale factor $a(t)$, with 
$t$ the cosmic time. 
Such a field could be the
dilaton, or other modulus field from the string multiplet~\cite{strings}.

{} Consider the lowest order 
Friedmann equation, as well as the 
equation of motion of the field $\phi$ in $D+1$ dimensions (the overdot 
denotes cosmic-time derivative), which are (formally) derived from 
the $\sigma$-model $\beta$-functions of a perturbative string theory
\bea
&& H^2 \equiv \left(\frac{\dot a}{a}\right) = \frac{2\kappa^2}{D(D-1)}E 
=  \frac{({\dot \phi})^2}{2} + V(\phi)~, \nonumber \\
&& {\ddot \phi} + DH{\dot \phi} + V'(\phi) = 0~, 
\eea
with $E$ the total energy of the scalar field, $V$ its potential, 
and a prime indicating variations with respect to the field $\phi$.
We obtain the following 
expressions
for the scale factor $a(t)$ 
and the cosmic horizon $\delta $: 
\bea
a(t)  &=& {\rm exp}\left(\int d\phi \sqrt{\frac{E}{D(D-1)(E-V)}}\right)~, \nonumber \\
\delta & = & \int^\infty \frac{dt}{a} = \int d\phi \frac{1}{a{\dot \phi}}
= \int d\phi \frac{1}{a\sqrt{2(E - V)}}
\label{conds}
\eea
The condition for the existence of a cosmic horizon is of course the 
convergence of the integral on the right-hand-side of 
the expression for $\delta$. This depends on the asymptotic behaviour
of the potential $V$ as compared to the total energy $E$.  
This behaviour can be studied in a generic perturbative 
string theory, based on the form of low energy potentials 
of possible quintessence candidates, such as dilaton, moduli etc.
Because realistic string theories involve at a certain stage 
supersymmetry in target space, which is broken as we 
go down to the four dimensional world after compactification, or as 
we lower the energy from the string (Planck) scale, 
such arguments depend on the form of the potential, dictated
by supersymmetry-breaking considerations. The form is such that 
$\delta \to \infty$ in (\ref{conds}), and hence there are 
no horizons.  I will not repeat these arguments here, because 
the above-mentioned
CPT/scattering-matrix based argument is more general, and
encompasses such cases, and is the most fundamental reason
for incompatibility of perturbative strings with space-time 
backgrounds with horizons. I refer the interested reader to the 
literature~\cite{banks}.  

I would like to stress, however, that these 
arguments refer to the traditional critical strings, without branes,
where a low-energy field theory derives from conformal 
invariance conditions. 
From this latter point of view it is straightforward to 
understand the problem of incorporating cosmologies with 
horizons, such as inflation or in general de Sitter space times, 
in perturbative strings. A tree-world-sheet $\sigma$-model
on, say, graviton backgrounds, whose conformal 
invariance conditions would normally yield
the target-space geometry,  reads to order $\alpha '$ ($\alpha '$ denotes henceforth the Regge slope)~\cite{strings}:
\be
\beta_{\mu\nu} = R_{\mu\nu} + \dots 
\label{beta} 
\ee
where the $\dots$ indicate contributions 
from other background fields, such as dilaton {\it etc.}.

Ignoring the other fields, conformal invariance of the pertrurbative
stringy $\sigma$-model would require a Ricci flat $R_{\mu\nu} = 0$ background,
which is not the case of a dS space, for which (c.f. (\ref{einsteq})) 
\be
R_{\mu\nu} = \Lambda g_{\mu\nu} 
\label{dscontr}
\ee
To generate such corrections in the early days of string theory,
Fischler and Susskind~\cite{sussk} 
had to invoke renormalization-group corrections to the above-tree level 
$\beta$-function (\ref{beta}), 
induced by higher string loops, i.e. higher topologies of the 
$\sigma$-model world sheet. Tadpoles ${\cal J}$ 
of dilatons at one string loop order
(torus topologies) yielded a dS (or AdS depending on the sign of ${\cal J}$) 
type contribution to the graviton $\beta$-function, ${\cal J}g_{\mu\nu}$.
The basic idea behind this approach is to accept that world-sheet 
surfaces of higher topologies with handles whose size is smaller than the 
short-distance cutoff of the world-sheet theory, 
will not be `seen' as higher- topologies but appear `effectively' as tree 
level ones. They will, therefore, lead to loop corrections to the 
traditional tree-level $\beta$-functions of the various background fields,
which cannot be discovered at tree level. Conformal invariance implies of 
course that tori with such small handles are equivalent to world-sheet spheres 
but with a long thin tube connected to them. For more details on this 
I refer the interested reader in my lectures 
in the first Aegean School~\cite{cosmology}. 

Nevertheless, this approach does not solve the problem, despite its formal 
simplicity and elegance. The reason is two fold: first, string-loop 
perturbation theory is not Borel-resummable, and as such, the expansion in 
powers of genus of closed Riemann surfaces with handles 
(and holes if open strings are included), does not converge mathematically,
hence it cannot give sensible answers for strong or intermediate string 
couplings. It is indeed, expected, that the dark energy  is a property of 
a full theory of quantum gravity, and as such, an explanation 
of it should {\it not} be restricted only to perturbative string theory.
Second, a string propagating in a space-time with a loop-induced cosmological
constant will not be characterised by a well-defined 
scattering matrix, which by definition,
as already mentioned, is a `must' for perturbative string theory.

Thus, the issue remains as to what kind of dark energy one is likely
to encounter in string theory.

\section{Dilaton quintessence and String Theory}

\subsection{An expanding Universe in String Theory} 

One of the simplest, and most natural quintessence fields, 
to generate a dynamical dark energy component for the 
string Universe is the dilaton, $\Phi$, a scalar field that appears
in the basic gravitational multiplet of any (super)string theory~\cite{strings}.
Dilaton cosmology has been originated by Antoniadis, Bachas, Ellis and 
Nanopoulos in \cite{aben}, where the 
basic steps for a correct formulation of an expanding Robertson-Walker 
Universe in string theory have been taken, consistent with conformal invariance
conditions~\footnote{In fact that work 
was actually the first work on Liouville 
supercritical strings~\cite{ddk}, with the Liouville mode identified with the target time, although this had not been recognised in the original work, 
but later~\cite{emn}.} .
The crucial r\^ole of a time dependent dilaton 
field had been emphasized. 

In~\cite{aben} 
a time-dependent dilaton background, with a linear
dependence on time in the so-called $\sigma$-model frame was assumed. Such
backgrounds, even when the $\sigma$-model metric is flat, lead to exact
solutions (to all orders in $\alpha '$) of the conformal invariance
conditions of the pertinent stringy $\sigma$-model, and so are acceptable
solutions from a perturbative viewpoint. It was argued in~\cite{aben} that
such backgrounds describe linearly-expanding Robertson-Walker Universes,
which were shown to be exact conformal-invariant solutions, corresponding
to Wess-Zumino models on appropriate group manifolds.

The pertinent $\sigma$-model action in a background with graviton $G$,
antisymmetric tensor $B$ and dilaton $\Phi$ reads~\cite{strings}:
\begin{equation} S_\sigma = \frac{1}{4\pi\alpha '} \int_\Sigma d^2\xi
[\sqrt{-\gamma} G_{\mu\nu} \partial_\alpha X^\mu \partial^\alpha X^\nu +
i\epsilon^{\alpha\beta} B_{\mu\nu} \partial_\alpha X^\mu \partial_\beta
X^\nu + \alpha '\sqrt{-\gamma}R^{(2)}\Phi], \label{sigmamodel}
\end{equation} where $\Sigma$ denotes the world-sheet, with metric
$\gamma$ and the topology of a sphere, $\alpha$ are world-sheet indices,
and $\mu,\nu $ are target-space-time indices. The important point
of~\cite{aben} was the r\^ole of target time $t$ as a specific dilaton
background, linear in that coordinate, of the form
\begin{equation}
\label{lineardil}
\Phi = {\rm const} - \frac{1}{2}Q~t,
\end{equation}
where $Q$ is a constant and $Q^2 > 0$ is the $\sigma$-model central-charge
deficit, allowing this {\it supercritical} string theory to be formulated
in some number of dimensions different from the critical number.
Consistency of the underlying world-sheet conformal field
theory, as well as modular invariance of the string scattering amplitudes,
required {\it discrete} values of $Q^2$, when expressed in units of the
string length $M_s$~\cite{aben}.
This was the first example of a non-critical string cosmology, with the
spatial target-space coordinates $X^i$, $i=1, \dots D-1$, playing the
r\^ole of $\sigma$-model fields. This non-critical string was not
conformally invariant, and hence required Liouville dressing~\cite{ddk}.
The Liouville field had time-like signature in target space, since the
central charge deficit $Q^2 > 0$ in the model
of~\cite{aben}, and its zero mode played the r\^ole of target time.

As a result of the non-trivial dilaton field, the 
Einstein term in the effective $D$-dimensional low-energy
field theory action is conformally rescaled by $e^{-2\Phi}$.
This requires a redefinition 
of the $\sigma$-model-frame space-time metric $g_{\mu\nu}^\sigma$ to
the `physical' Einstein metric  $g_{\mu\nu}^E$:
\begin{equation}
g_{\mu\nu}^E = e^{-\frac{4\Phi}{D-2}}G_{\mu\nu}~.
\label{smodeinst}
\end{equation}
Target time must also be rescaled, so that the
metric acquires the standard Robertson-Walker (RW) form in the normalized
Einstein frame for the effective action:
\begin{equation}
ds^2_E = -dt_E^2 + a_E^2(t_E) \left(dr^2 + r^2 d\Omega^2 \right),
\end{equation}
where we show the example of a spatially-flat RW metric for definiteness,
and $a_E(t_E)$ is an appropriate scale factor, which is a function of $t_E$
alone in the homogeneous cosmological backgrounds we assume throughout.

The Einstein-frame time is related to the 
time in the $\sigma$-model frame~\cite{aben} by:
\begin{equation}\label{einsttime}
dt_E = e^{-2\Phi/(D-2)}dt \qquad \to \qquad t_E = \int ^t e^{-2\Phi(t')/(D-2)}
dt'~. 
\end{equation} 
The linear dilaton background (\ref{lineardil}) yields
the following relation between the Einstein and $\sigma$-model frame 
times:
\begin{equation} 
t_E = c_1 + \frac{D-2}{Q}e^{\frac{Q}{D-2}t},
\end{equation}
where $c_{1}$ is an appropriate (positive) constant.
Thus, a dilaton background (\ref{lineardil}) that is
linear in the $\sigma$-model time scales logarithmically with 
the Einstein time (Robertson-Walker cosmic time) $t_E$:
\begin{equation}\label{dil2}
\Phi (t_E) =({\rm const.}') - \frac{D-2}{2}{\rm ln}(\frac{Q}{D-2}t_E).
\end{equation} 
In this regime, the string coupling~\cite{strings}: 
\begin{equation}
g_s = {\rm exp}\left(\Phi(t)\right)
\label{defstringcoupl}
\end{equation}
varies with the cosmic time $t_E$ as 
$g_s^2 (t_E) \equiv e^{2\Phi} \propto \frac{1}{t_E^{D-2}}$,
thereby implying a vanishing effective string coupling 
asymptotically in cosmic time. 
In the linear dilaton background of~\cite{aben}, the asymptotic
space-time metric in the Einstein frame reads:
\begin{equation} \label{metricaben}
ds^2 = -dt_E^2 + a_0^2 t_E^2 \left(dr^2 + r^2 d\Omega^2 \right)
\end{equation}
where $a_0$ a constant.
Clearly, there is no acceleration in the expansion of the Universe 
(\ref{metricaben}).

The effective low-energy action on the four-dimensional 
brane world for the gravitational multiplet 
of the string in the Einstein frame reads~\cite{aben}:
\begin{equation}
S_{\rm eff}^{\rm brane} = \int d^4x\sqrt{-g}\{ R  - 2(\partial_\mu \Phi)^2 
- \frac{1}{2} e^{4\Phi}( \partial_\mu b)^2 - \frac{2}{3}e^{2\Phi}\delta c 
\},
\label{effaction}
\end{equation}
where $b$ is the four-dimensional axion field
associated with a four-dimensional representation of the antisymmetric
tensor, and $\delta c = C_{\rm int} - c^*$,
where $C_{\rm int}$ is the central charge of the conformal world-sheet
theory corresponding to the transverse (internal) string dimensions, and
$c^*=22 (6)$ is the critical value of this internal central charge of the
(super)string theory for flat four-dimensional space-times.  The linear
dilaton configuration (\ref{lineardil}) corresponds, in this language, to
a background charge $Q$ of the conformal theory, which contributes a term
$-3Q^2$ (in our normalization)  to the total central charge. The latter
includes the contributions from the four uncompactified dimensions of our
world.  In the case of a flat four-dimensional Minkowski space-time, one
has $C_{\rm total} = 4 -3Q^2 + C_{\rm int} = 4 - 3Q^2 + c^* + \delta c$,
which should equal 26 (10). This implies that $C_{\rm int} = 22 + 3Q^2~(6 
+
3Q^2)$ for bosonic (supersymmetric) strings.  

An important result in~\cite{aben} was the discovery of an exact conformal
field theory corresponding to the dilaton background (\ref{dil2}) and a
constant-curvature (Milne) static metric in the $\sigma$-model frame (or,
equivalently, a linearly-expanding Robertson-Walker Universe in the
Einstein frame).  The conformal field theory corresponds to a
Wess-Zumino-Witten two-dimensional world-sheet model on a group manifold
$O(3)$ with appropriate constant curvature, whose coordinates correspond
to the spatial components of the four-dimensional metric and antisymmetric
tensor fields, together with a free world-sheet field corresponding to the
target time coordinate. The total central charge in this more general case
reads $C_{\rm total} = 4 - 3Q^2 - \frac{6}{k+2}+ C_{\rm int}$, where $k$
is a positive integer corresponding to the level of the Kac-Moody algebra
associated with the WZW model on the group manifold. The value of $Q$ is
chosen in such a way that the overall central charge $c=26$ and the theory
is conformally invariant. Since such unitary conformal field theories have
{\it discrete} values of their central charges, which accumulate to
integers or half-integers from {\it below}, it follows that the values of
the central charge deficit $\delta c$ are {\it discrete} and {\it finite}
in number.  From a physical point of view, this implies that the
linear-dilaton Universe may either stay in such a state for ever, for a
given $\delta c$, or tunnel between the various discrete levels before
relaxing to a critical $\delta c =0$ theory. It was argued in~\cite{aben}
that, due to the above-mentioned finiteness of the set of allowed discrete
values of the central charge deficit $\delta c$, the Universe could reach
flat four-dimensional Minkowski space-time, and thus exit from the
expanding phase, after a finite number of phase transitions.

The analysis in~\cite{aben} also showed that there
are tachyonic mass shifts of order $-Q^2$ in the bosonic string
excitations, but not in the fermionic ones. This implies the appearance of
tachyonic instabilities and the breaking of target-space supersymmetry in
such backgrounds, as far as the excitation spectrum is concerned. The
instabilities could trigger the cosmological phase transitions, since they
correspond to relevant renormalization-group world-sheet operators, and
hence initiate the flow of the internal unitary conformal field theory
towards minimization of its central charge, in accordance with the
Zamolodchikov $c$-theorem~\cite{zam}. In
semi-realistic cosmological models~\cite{dgmpp} such tachyons decouple
from the spectrum relatively quickly. On the other hand, as a result of
the form of the dilaton in the Einstein frame (\ref{dil2}), we observe
that the dark-energy density for this (four-dimensional)  Universe,
$\Lambda \equiv e^{2\Phi}\delta c$, is relaxing to zero with a
$1/t_E^{(D-2)}$ dependence on the Einstein-frame time for each of the
equilibrium values of $\delta c$.  Therefore, the breaking of
supersymmetry induced by the linear dilaton is only an
obstruction~\cite{witten}, rather than a spontaneous breaking, in the
sense that it appears only temporarily in the boson-fermion mass
splittings between the excitations, whilst the vacuum energy of the
asymptotic equilibrium theory vanishes.

\subsection{Pre Big Bang Scenaria}

\begin{figure}[t]
\centering
\includegraphics[width=8.0cm]{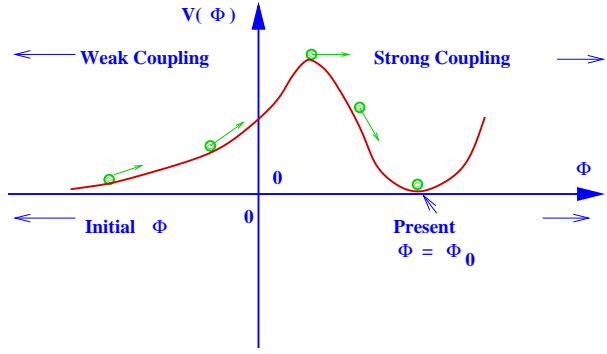} 
\caption { The dilaton potential in the pre Big-Bang scenario
of string cosmology. The string coupling grows strong at late times,
and hence current-era is described by strongly-coupled strings, 
where higher string loop corrections matter.}
\label{fig:preBB}
\end{figure}


After the work of \cite{aben}, dilaton cosmology has been 
discussed in a plethora of interesting works, most of them 
associated with the so-called `pre-Big-Bang' (pBB) 
cosmologies~\cite{veneziano}, 
suggested by Veneziano, and pursued further by Gasperini, Veneziano 
and collaborators.
For the interested reader, this type of cosmology has been reviewed 
by the author
in the first Aegean School~\cite{cosmology}, where I refer the 
interested reader for more details. 

The basic feature behind the approach, is the fact that 
the dilaton has such time dependence in these models that, as the cosmic time elapses, the string coupling $g_s = e^\Phi$ grows stronger at late stages of the Universe. The dilaton potential in the pre Big-Bang approach, which may be generated by higher string loops, has the generic form depicted 
in fig.~\ref{fig:preBB}~\cite{veneziano}. The situation is opposite that of 
\cite{aben}, where as we have seen 
the string coupling becomes weaker 
with the cosmic time, and perturbative strings 
are sufficient for a description of the Universe at late epochs. 

I will not discuss in detail the pBB theories, 
since there are excellent reviews on the subject~\cite{veneziano}, where 
I refer the interested reader for more details. 
For our purposes here, I would like to emphasize 
the basic predictions of this model regarding the r\^ole of dilaton as a 
a quintessence field, responsible for 
late-time acceleration.

The starting point is the string-frame, low-energy, string-inspired 
effective action with 
graviton and dilaton backgrounds~\cite{strings}, 
to lowest order in the $\a'$ expansion, but
including dilaton-dependent loop (and non-perturbative) corrections, 
which are essential given that at late epochs the dilaton grows strong
in pBB scenaria. Such corrections are 
encoded in a few  ``form factors"~\cite{veneziano}  
$\psi(\phi)$,  $Z(\phi)$,
$\alpha{(\phi)}$, $\dots$, and in an effective dilaton potential $V(\phi)$.
The effective action reads: 
\bea
S &=& -{M_s^{2}\over 2} \int d^{4}x \sqrt{- \wt g}~
\left[e^{-\psi(\phi)}\widetilde R+ Z(\phi)
\left(\wt\nabla \phi\right)^2 + {2\over M_s^{2}} V(\phi)\right] 
\nonumber \\
&-& {1\over 16 \pi} \int d^{4}x {\sqrt{- \wt g}~  \over  \alpha{(\phi)}}
F_{{\mu\nu}}^{2} + \Ga_{m} (\phi, \wt g, \rm{matter})  
\label{3}
\eea
where we follow the conventions of \cite{veneziano}.

The four dimensional action above is the result 
of compactification. It is also assumed that  
the the corresponding moduli 
have been frozen 
at the string scale. 
In the approach of \cite{veneziano} 
it is assumed  that the
form factors $\psi(\phi)$,  $Z(\phi)$, $\alpha{(\phi)}$ approach a finite
limit as $\phi \rightarrow + \infty$ while, in the
same limit,  $V \rightarrow 0$. 
The fields appearing in the matter action $\Ga_{m}$ are
in general non-minimally and non-universally coupled to the dilaton 
(also because of the loop corrections).  

In the Einstein frame the action (\ref{3}) 
becomes
\bea 
S &=& -{M_P^{2}\over 2}\int d^{4}x \sqrt{- g}~
\left[ R - {k(\phi)^{2}\over 2} 
\left(\nabla \phi\right)^2  + { 2\over M_P^{2}}\hat{V}(\phi)\right]
\nonumber\\ 
&-&   {1\over 16 \pi} \int d^{4}x {\sqrt{- g}~  \over  \alpha{(\phi)}}
F_{{\mu\nu}}^{2} + \Ga_{m} (\phi, c_1^2 g_{\mu\nu}\epsi, \rm{matter})
\;,  
\label{EFaction}
\eea
where 
\be
k^2(\phi) = 3 \psi^{\prime 2} - 2 \epsi Z , 
~~~~~~~~~~~~~\hat V = c_1^4 e^{2\psi} 
V \; .
\ee
The pertinent equations of motion for the graviton field read: 
\bea
&&
6H^2 = \r +\r_\phi, \label{11} \\
&&
4 \dot H + 6H^2 =-p -p_\phi,
\label{12}
\eea
while the dilaton equation is: 
\be
k^2(\phi) \left(\ddot{\phi}+3 H \dot{\phi}\right) +
k(\phi)\, k'(\phi)\, \dot{\phi}^2 
+ \hat{V}'(\phi) + \frac{1}{2}\left[{\psi'(\phi)} (\rho - 3 p) + \sg
\right] = 0.
\label{13}
\ee
In the above equations $H= \dot a /a$, a dot denotes differentiation
with respect to the Einstein cosmic time, and we have used the
definitions:
\be
\r_\phi= {1\over2} k^2(\phi) \dot \phi^2 +\hat V(\phi), ~~~~~~~
p_\phi= {1\over2} k^2(\phi) \dot \phi^2 -\hat V(\phi). 
\label {14}
\ee
After some manipulations the pertinent equations of motion,
describing the dynamics of the system, read: 
\bea
&&
2\, H^2 \, k^2 \, \ \frac{d^2\phi}{d\chi^2}\, +\, k^2 \left(\frac{1}{2}\rhm
+
\frac{1}{3}\r_r
+ \hat{V}\right)\frac{d \phi}{d\chi} \, +\, 
2 H^2\, k\, k'  \left(\frac{d\phi}{d\chi}\right)^2 + \nonumber \\
&& 2 \hat{V}' + \psi' \rhm \
+ \sg = 0~, \nonumber \\
&&
 H^2 \left[6 -\frac{k^2}{2}\left(\frac{d\phi}{d\chi}\right)^2 \right] \,
\ =\ \rhm + \r_r + \hat{V}.
\label{19}
\eea
where $\chi = {\rm ln} a$, with 
$a$ the scale factor in units of the present day scale. 

The matter evolution equation, on the other hand, 
can be split into the various
components (radiation (r), baryonic (d)): 
\bea
&&
\frac{d \r_r}{d\chi} + 4 \r_r - \frac{\sg_r}{2} \frac{d \phi}{d\chi}  = 0,
\nonumber \\
&&
\frac{d \r_b}{d\chi} + 3 \r_b -\frac{1}{2}\left( \psi' \r_b + \sg_b \right)
\frac{d\phi}{d\chi}\ = \ 0.
\nonumber \\
&&
\frac{d \r_d}{d\chi} + 3 \r_d -\frac{1}{2}\left( \psi' \r_d + \sg_d \right)
\frac{d\phi}{d\chi}\ = \ 0.
\label{rhodark}
\eea
And for the dilaton-energy density $\rho_\phi$ one can obtain the equation 
\be
\frac{d \rho_\phi}{d\chi} + 6 \rho_\phi - 6 \hat{V}(\phi) + \frac{1}{2}
\left( \psi' \r_m + \sg \right) \frac{d\phi}{d\chi} \ = \ 0 .
\label{22}
\ee
The analysis of \cite{veneziano}, based on these equations, 
leads to predictions regarding 
the behaviour of the various 
cosmological parameters of the pBB dilaton cosmology.

Under various approximations and assumptions, which I will not go through,
but I would stress that they are due to the fact that the 
various form factors and the dilaton couplings to matter are not known
in this approach due to the (uncontrolled) loop corrections, 
one can obtain the asymptotic
evolution of the Hubble factor and of the dominant energy density in this 
approach,
\be
H \sim a^{-3/(2+q)}, ~~~~~~~~~~~~~~
\r \sim a^{-6/(2+q)}. 
\label{318}
\ee
where $q = {\cal O}(1)$ and 
is expressed in terms of the 
various energy densities in the model  
$q = q = 2\, \frac{\Omega_V - \Omega_k}{1 + \Omega_k - \Omega_V}$. 

\begin{figure}[t] 
\begin{center}
\includegraphics[width=4.0cm]{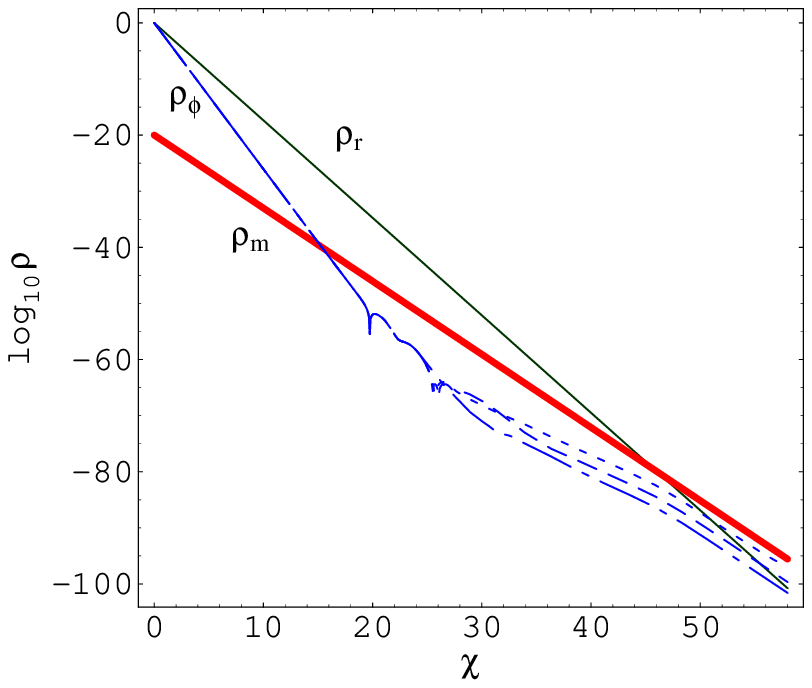}
\includegraphics[width=4.0cm]{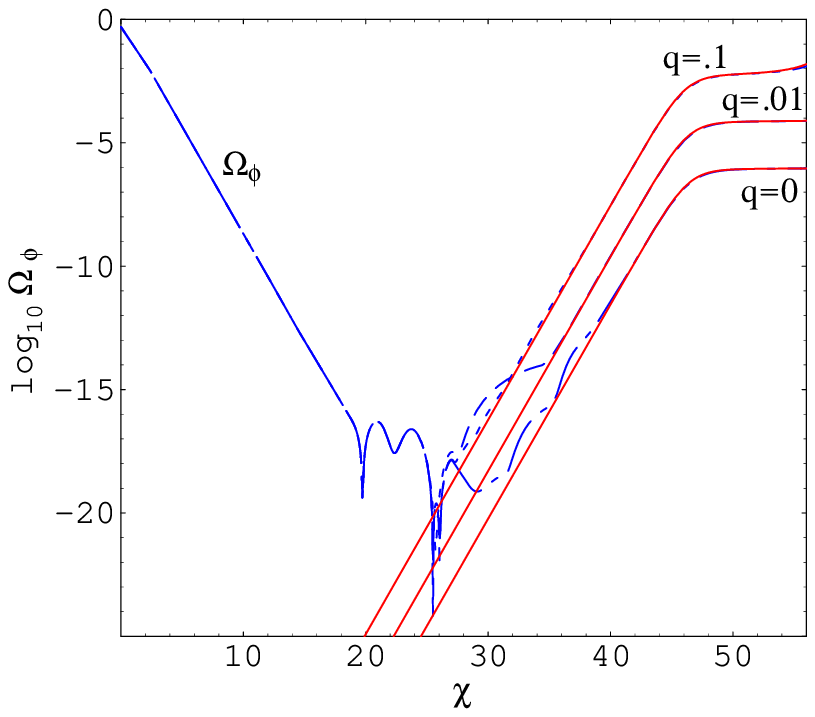}
\vskip 5mm
\caption[] {Time evolution of $\r_\phi$ for $q=0$
(dash-dotted curve), $q=0.01$ (dashed curve) and $q=0.1$ (dotted
curve). The initial scale is $a_i= 10^{-20} a_{\rm eq}$, and 
the epoch of matter-radiation
equality corresponds
to $\chi \simeq 46$. Left panel: the dilaton energy density is compared
with the radiation (thin solid curve) and matter (bold solid curve) energy
density. Right panel: the dilaton energy density (in critical units) is
compared with the analytical estimates
for the focusing and dragging phases.}
\end{center}
\label{fig:gasp1}
\end{figure}

\begin{figure}[t]
\begin{center}
\includegraphics[width=4.0cm]{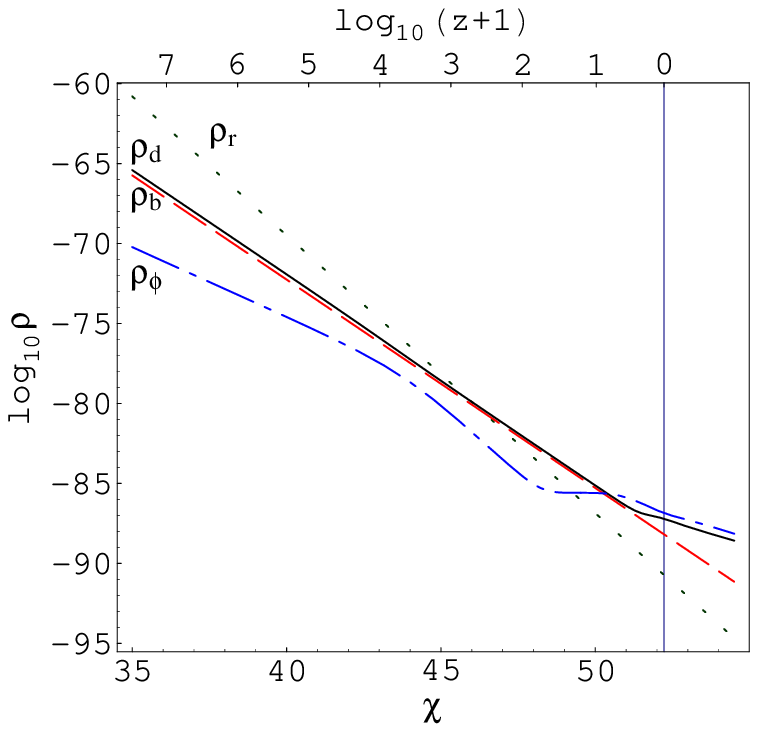}
\includegraphics[width=4.0cm]{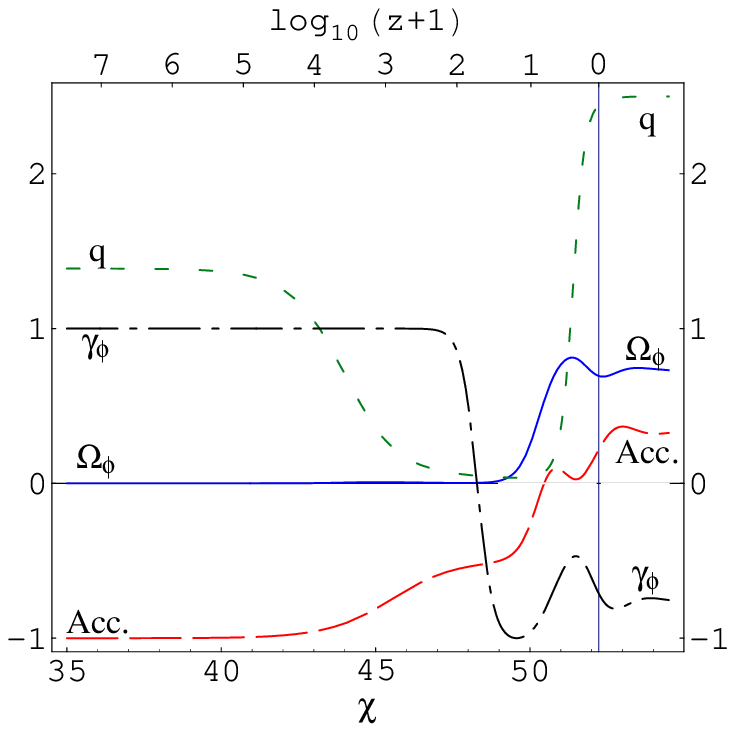}
\vskip 5mm
\caption[] {Left panel: Late-time evolution of
the dark matter (solid curve), baryonic matter (dashed curve), radiation
(dotted curve) and the dilaton (dash-dotted curve) energy densities,
for the pBB string cosmology model of \cite{veneziano}.
The upper horizontal axis gives the $\log_{10}$ of the redshift
parameter.  Right panel: for the same model, the late-time evolution of 
$q$ (fine-dashed curve), $w_\phi$ (dash-dotted curve),  
$\Omega_\phi$ (solid curve) and of the acceleration
parameter $\ddot{a}a/\dot{a}^2$ (dashed curve).}
\end{center}
\label{fig:gasp2} 
\end{figure}

The evolution of the various cosmological parameters in a typical of such 
pBB models is given in figures 4 and 5,  
taken from the second ref. in \cite{veneziano}.

\subsection{Non Critical Strings and Dark Energy} 

Pre Big Bang scenaria, as we have discussed, involve strong string 
couplings, and hence the various form factors appearing in the 
effective actions are unknown. 

An alternative approach, is to invoke the weak coupling late-era  dilaton 
cosmology of \cite{aben}, which 
has the advantage that at late eras perturbative 
$\sigma$-model calculations are reliable, and hence one can 
perform concrete computations and predictions. 
The analysis of \cite{aben} however has to be generalised to 
include inflationary and other backgrounds with horizons, 
if the dark matter issue and accelerating Universes are to be tackled. 
This cannot be 
achieved with the simple linear dilaton backgrounds of \cite{aben}.

In~\cite{emn} we went one step beyond the analysis in~\cite{aben}, and
considered more complicated $\sigma$-model metric backgrounds that did not
satisfy the $\sigma$-model conformal-invariance conditions, and therefore
needed Liouville dressing~\cite{ddk} to restore conformal invariance. Such
backgrounds could even be time-dependent, living in $(d+1)$-dimensional
target space-times. Various mathematically-consistent forms of
non-criticality can be considered, for instance cosmic catastrophes such
as the collision of brane worlds~\cite{gravanis,brany}. Such models lead
to supercriticality of the associated $\sigma$ models describing stringy
excitations on the brane worlds.  The Liouville dressing of such
non-critical models results in $(d+2)$-dimensional target spaces with two
time directions.  An important point in~\cite{emn} was the identification
of the (world-sheet zero mode of the) Liouville field with the target
time, thereby restricting the Liouville-dressed $\sigma$ model to a
$(d+1)$-dimensional hypersurface of the $(d+2)$-dimensional target space, 
thus maintaining the initial target space-time dimensionality. We
stress that this identification is possible only in cases where
the initial $\sigma$ model is supercritical, so that the Liouville mode
has time-like signature~\cite{aben,ddk}. In certain
models~\cite{gravanis,brany}, such an identification was proven to be
energetically preferable from a target-space viewpoint, since it minimized
certain effective potentials in the low-energy field theory corresponding
to the string theory at hand.

\begin{figure}[tb]
\begin{center}
\includegraphics[width=5cm]{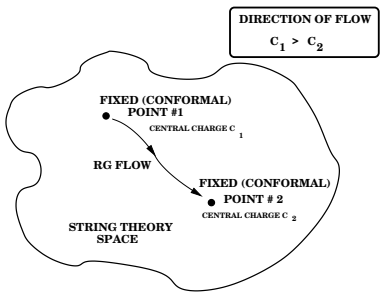}
\end{center}
\caption{\it A schematic view of string 
theory space, which is an infinite-dimensional
manifold endowed with a (Zamolodchikov) metric. The dots denote conformal
string backgrounds. A non-conformal string flows (in a two-dimensional
renormalization-group sense)  from one fixed point to another, either of
which could be a hypersurface in theory space.  The direction of the flow
is irreversible, and is directed towards the fixed point with a lesser
value of the central charge, for unitary theories, or, for general
theories, towards minimization of the degrees of freedom of the system.}
\label{fig:flow}
\end{figure}

All such cosmologies require some physical reason for the initial
departure from the conformal invariance of the underlying $\sigma$ model
that describes string excitations in such Universes. The reason could be
an initial quantum fluctuation, or, in brane models, a catastrophic cosmic
event such as the collision of two or more brane worlds.  Such
non-critical $\sigma$ models relax asymptotically to conformal $\sigma$
models, which may be viewed as equilibrium points in string theory space,
as illustrated in Fig.~\ref{fig:flow}. In some interesting cases of
relevance to cosmology~\cite{dgmpp}, which are particularly generic, the
asymptotic conformal field theory is that of~\cite{aben} with a linear
dilaton and a flat Minkowski target-space metric in the $\sigma$-model
frame. In others, the asymptotic theory is characterized by a constant
dilaton and a Minkowskian space-time~\cite{gravanis}.
Since, as we discussed in \cite{emn} and review briefly 
below, the evolution of the central-charge deficit of
such a non-critical $\sigma$ model, $Q^2(t)$, plays a crucial r\^ole in
inducing the various phases of the Universe, including an inflationary
phase, graceful exit from it, thermalization and a contemporary phase of
accelerating expansion, we term such Liouville-string-based cosmologies 
{\it Q-Cosmologies}.

The use of Liouville strings to describe the evolution of our Universe has
a broad motivation, since non-critical strings are associated with
non-equilibrium situations, as are likely to have occurred in the early
Universe. The space of non-critical string theories is much larger than
that of critical strings. It is therefore remarkable that the departure
from criticality may enhance the predictability of string theory to the
extent that a purely stringy quantity such as the string coupling $g_s$
may become accessible to experiment via its relation to the present-era
cosmic acceleration parameter: $g_s^2 = -q^{0}$~\cite{emn04}.  Another
example arises in a non-critical string approach to inflation, if the Big
Bang is identified with the collision of two
D-branes~\cite{brany}. In such a scenario, astrophysical observations may
place important bounds on the recoil velocity of the brane worlds after
the collision, and lead to an estimate of the separation of the branes at
the end of the inflationary period.

In such a framework, the identification of target time with a world-sheet
renormalization-group scale, the zero mode of the Liouville
field~\cite{emn}, provides a novel way of selecting the ground state of
the string theory. This is not necessarily associated with minimization of
energy, but could simply be a result of cosmic chance. It may be a random
global event that the initial state of our cosmos corresponds to a certain
Gaussian fixed point in the space of string theories, which is then
perturbed into a Big Bang by some relevant (in a world-sheet sense)
deformation, which makes the theory non-critical, and hence out of
equilibrium from a target space-time viewpoint. The theory then flows, as
indicated in Fig.~\ref{fig:flow}, along some specific
renormalization-group trajectory, heading asymptotically to some ground
state that is a local extremum
corresponding to an infrared fixed point of this perturbed world-sheet 
$\sigma$-model theory. This approach allows for many `parallel
universes' to be implemented, and our world might be just one of these.
Each Universe may flow between different fixed points, its trajectory
following a perturbation by a different operator.  It seems to us that
this scenario is more attractive and specific than the landscape
scenario~\cite{sussk}, which has recently been advocated as a framework
for parametrizing our ignorance of the true nature of string/M theory.

Let us briefly review the basic formalism. 
We consider a $\sigma$-model action deformed by 
a family of 
vertex operators $V_i$, corresponding to 
`couplings' $g^i$, which represent \emph{non-conformal} background space-time
fields from the massless string multiplet, such as 
gravitons, $G_{\mu\nu}$, antisymmetric tensors, $B_{\mu\nu}$,
dilatons $\Phi$, their supersymmetric partners, \emph{etc.}:
\begin{equation}
S=S_{0}\left( X\right) +\sum_{i}g^{i}\int d^{2}z\,V_{i}\left( 
X \right)~,  
\label{action}
\end{equation}
where $S_0$ represents a conformal $\sigma$ model describing 
an equilibrium situation.
The non-conformality of the background means 
that the pertinent $\beta^i$ function $\beta^i \equiv dg^i/d{\rm ln}\mu 
\ne 0$, where $\mu$ is a world-sheet renormalization scale. 
Conformal invariance would imply restrictions on the 
background fields/$\sigma$-model couplings, $g^i$, corresponding to 
the constraints $\beta^i = 0$, which are equivalent to equations
of motion derived from a target-space effective action for the corresponding
fields $g^i$.  The entire low-energy 
phenomenology and model building of critical string theory is based on such
restrictions~\cite{strings}.

In the non-conformal case $\beta^i \ne 0$, the theory is in need of
dressing by the Liouville field $\phi$ in order to restore conformal
symmetry~\cite{ddk}. The field $\phi $ acquires dynamics through the
integration over world-sheet covariant metrics in the path integral, and
may be viewed as a local dynamical scale on the world sheet~\cite{emn}.  
If the central charge of the (supersymmetric)  matter theory is $c_{m}>25
(9)$ (i.e., supercritical), the signature of the kinetic term of the
Liouville coordinate in the dressed $\sigma$-model is opposite to that of
the $\sigma$-model fields corresponding to the other target-space
coordinates.  As mentioned previously, this opens the way to the important
step of interpreting the Liouville field physically by identifying its
world-sheet zero mode $\phi _{0}$ with the target time in supercritical
theories~\cite{emn}.  Such an identification emerges naturally from the
dynamics of the target-space low-energy effective theory by minimizing the
effective potential~\cite{gravanis}.

In terms of the Liouville renormalization-group scale, one
has the following equation relating Liouville-dressed
couplings $g^i$ and $\beta$ functions in the non-critical string case:
\begin{equation}
{\ddot  g}^i + Q{\dot g}i = \mp\beta^i(g_j)~,
\label{liouvilleeq}
\end{equation}
where the - (+) sign in front of the $\beta$-functions 
on the right-hand-side applies to super(sub)critical strings, 
the overdot denotes differentiation with respect to the 
Liouville zero mode, $\beta^i$ is the world-sheet renormalization-group
$\beta$ function (but with the renormalized couplings replaced 
by the Liouville-dressed ones as defined by the procedure in \cite{ddk}), 
and the minus sign on the right-hand side (r.h.s.) of (\ref{liouvilleeq}) 
is due to the time-like signature of the Liouville field.
Formally, the $\beta^i$ of the r.h.s.\ of (\ref{liouvilleeq})
may be viewed as 
power series in the (weak) couplings $g^i$. 
The covariant (in theory space)
${\cal G}_{ij}\beta^j$ function,
with ${\cal G}_{ij}$ the (Zamolodchikov) metric,
 may be expanded as:
\begin{equation}
{\cal G}_{ij}\beta^j = 
\sum_{i_n} \langle V_i^L V_{i_1}^L \dots V_{i_n}^L \rangle_\phi g^{i_1} 
\dots g^{i_n}~,
\label{betaexp}
\end{equation}
where $V_i^L$ indicates Liouville dressing  as in \cite{ddk}
$ \langle \dots \rangle_\phi = \int d\phi d\vec{r}~{\rm exp}(-S(\phi, \vec{r}, g^j))$
denotes a functional average including Liouville integration, and 
$S(\phi, \vec{r}, g^i)$ is the Liouville-dressed $\sigma$-model
action, including the Liouville action~\cite{ddk}.

In the case of stringy $\sigma$ models, 
the diffeomorphism invariance of the target space results in the 
replacement of (\ref{liouvilleeq}) by: 
\begin{equation} 
  {\ddot g}^i + Q(t){\dot g}^i = \mp{\tilde \beta}^i ,
\label{liouvilleeq2}
\end{equation} 
where the ${\tilde \beta}^i$ are the Weyl anomaly coefficients of the 
stringy $\sigma$ model in the background $\{ g^i \}$, which differ 
from the ordinary world-sheet renormalization-group $\beta^i$ functions
by terms of the form:
\begin{equation}
{\tilde \beta}^i = \beta^i + \delta g^i 
\end{equation}
where $\delta g^i$ denote transformations of the 
background field $g^i$ under infinitesimal general coordinate
transformations, e.g., for gravitons~\cite{strings} 
${\tilde \beta}^G_{\mu\nu} = 
\beta^G_{\mu\nu} + \nabla_{(\mu} W_{\nu)}$, 
with $W_\mu = \nabla _\mu \Phi$,
and $\beta^G_{\mu\nu} = R_{\mu\nu}$ to order $\alpha '$ 
(one $\sigma$-model loop).

The set of equations (\ref{liouvilleeq}),(\ref{liouvilleeq2})  defines the
\emph{generalized conformal invariance conditions}, expressing the
restoration of conformal invariance by the Liouville mode.  The solution
of these equations, upon the identification of the Liouville zero mode
with the original target time, leads to constraints in the space-time
backgrounds~\cite{emn,gravanis}, in much the same way as the conformal
invariance conditions $\beta^i = 0$ define consistent space-time
backgrounds for critical strings~\cite{strings}.  It is important to
remark~\cite{emn} that the equations (\ref{liouvilleeq2}) can be
derived from the {\it variation} 
of {\it an off shell} 
action.  This follows from general properties of the
Liouville renormalization group, which guarantee that the appropriate
Helmholtz conditions in the string-theory space $\{ g^i \}$ for the
Liouville-flow dynamics to be derivable from an action principle are
satisfied. 

When applied to dilaton cosmologies, with dilaton and graviton 
backgrounds,  
this approach yields interesting results, 
including a modified asymptotic scaling of the dark matter energy density, 
$a^{-2}$ with the scale
factor, 
as well as 
an expression of the current-era acceleration parameter of the Universe
roughly proportional to the square of the string coupling, $q_0 \propto -(g_s^{0})^2$,
$g_s^2 = e^{2\Phi}$,
with $\Phi$ the current era dilaton (this proportionality relation 
becomes exact at late eras, when the matter 
contributions become negligible due to cosmic dilution).
The current-era dark energy in this framework relaxes to zero with the 
Einstein cosmic time as $1/t^2$, and this scaling law follows from the 
generalised conformal invariance conditions (\ref{liouvilleeq2}), 
characterising the Liouville theory, as well as the identification 
of time with the 
Liouville mode~\cite{emn}. 

To be specific, after this identification, 
the relevant Liouville equations (\ref{liouvilleeq2}) 
for dilaton and graviton cosmological  
backgrounds, in the Einstein frame \cite{aben}, 
read~\cite{emn04}: 
\begin{eqnarray} 
&&3 \; H^2 - {\tilde{\varrho}}_m - \varrho_{\phi}\;=\; \frac{e^{2 \phi}}{2} \; \tilde{\cal{G}}_{\phi} \nonumber  \\
&&2\;\dot{H}+{\tilde{\varrho}}_m + \varrho_{\phi}+
{\tilde{p}}_m +p_{\phi}\;=\; \frac{\tilde{\cal{G}}_{ii}}{a^2} \nonumber  \\
&& \ddot{\phi}+3 H \dot{\phi}+ \frac{1}{4} \; \frac{\partial { \hat{V}}_{all} }{\partial \phi}
+ \frac{1}{2} \;( {\tilde{\varrho}}_m - 3 {\tilde{p}}_m )= 
- \frac{3}{2}\; \frac{ \;\tilde{\cal{G}}_{ii}}{ \;a^2}- \,
\frac{e^{2 \phi}}{2} \; \tilde{\cal{G}}_{\phi}  \; .\label{eqall}
\end{eqnarray} 
where $\tilde \rho _m$ and $\tilde p_m$ denote the matter 
energy density and pressure respectively, including dark matter contributions. 
As usual, the overdot denotes derivatives with 
respect to the Einstein time, 
and $H$ is the Hubble parameter of the Robertson-Walker Universe. 
The r.h.s of the above equations
denotes the non-critical
string {\it off-shell} terms, due to the non-equilibrium 
nature of the pertinent cosmology. The latter could be due to 
an initial cosmically catastrophic event, such as the 
collision of two brane worlds: 
\begin{eqnarray} && \tilde{\cal{G}}_{\phi} \;=\; e^{\;-2 \phi}\;( \ddot{\phi} -
{\dot{\phi}}^2 + Q e^{\phi} \dot{\phi}) \nonumber \\ &&
\tilde{\cal{G}}_{ii} \;=\; 2 \;a^2 \;(\; \ddot{\phi} + 3 H \dot{\phi}
+ {\dot{\phi}}^2 + ( 1 - q ) H^2 + Q e^{\phi} ( \dot{\phi}+ H )\;) \;
.  \end{eqnarray} 
Notice the {\it dissipative} terms proportional 
to $Q\dot{\phi}$, which are responsible for the terminology
``Dissipative Cosmology'' used alternatively for Q-cosmology~\cite{emn04}. 
In these equations, $q$ is the deceleration $q \equiv - \ddot{a} a /
{\dot{a}}^2$.  
The potential appearing in (\ref{eqall})
is defined by 
$\; {\hat{V}}_{all}= 2 Q^{\;2} \exp{\;( 2 \phi )}+V \;$ where, for the 
sake of generality, 
we have allowed 
for an additional  potential term in the string action $- \sqrt{-G}\;V$.

\begin{figure}[ht] 
\begin{center}
\includegraphics[width=4cm]{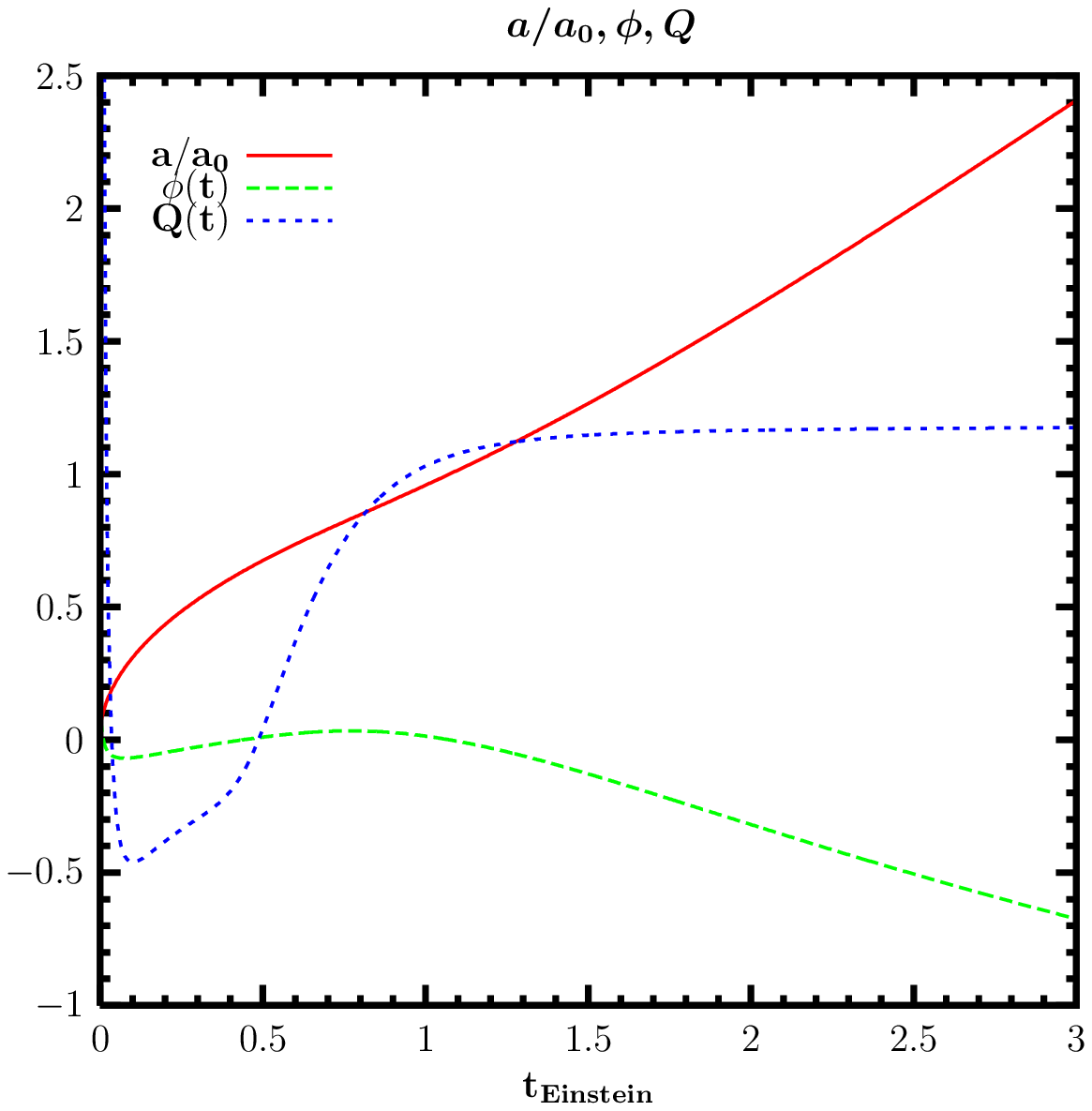}
\includegraphics[width=4cm]{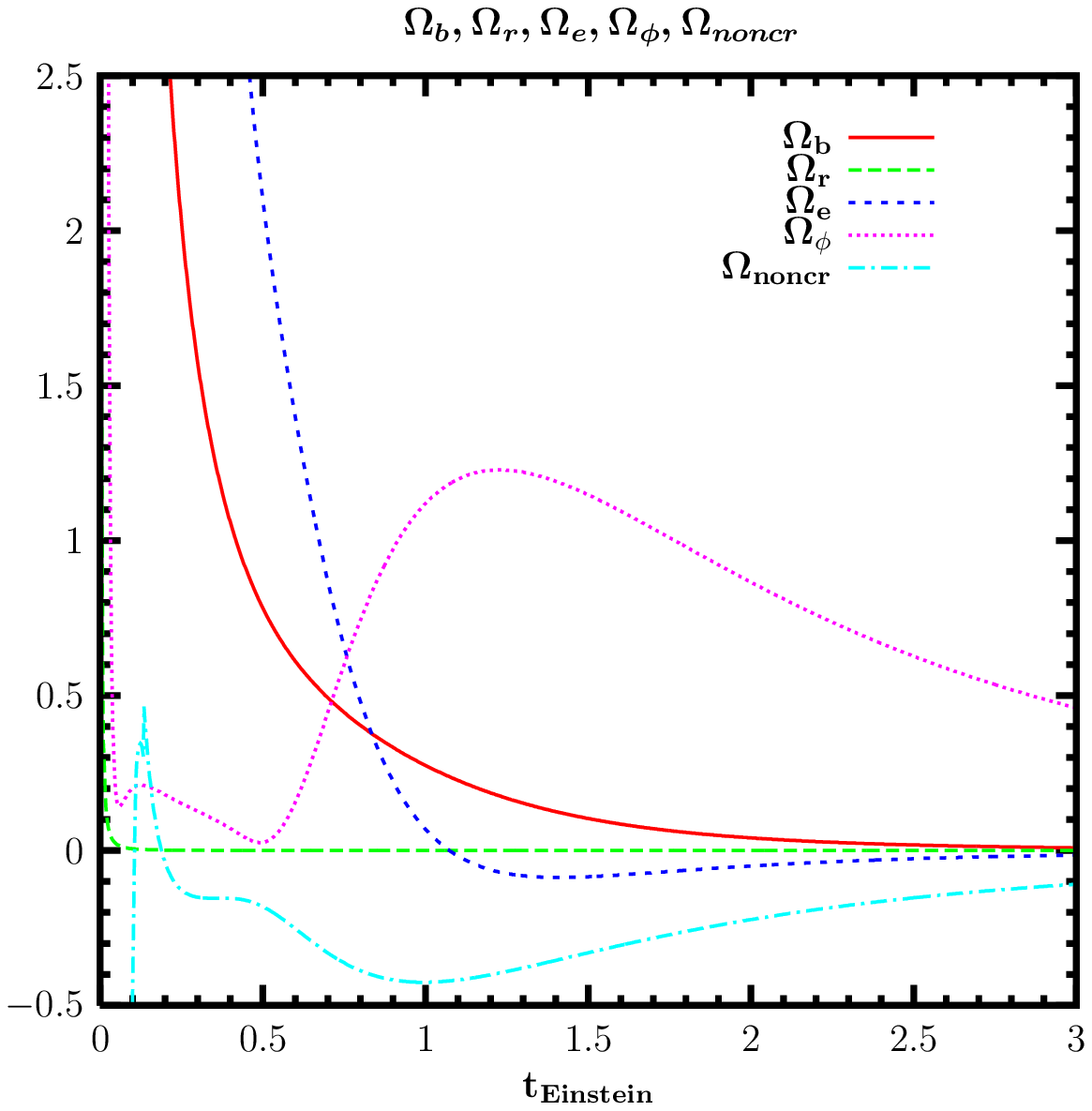}
\end{center}
\caption[]{ Left panel: The dilaton $\phi$, the (square root of the) 
central charge deficit
$Q$ and the ratio $a/a_0$ of the cosmic scale factor as functions of
the Einstein time $t_{Einstein}$. The present time is located where
$a/a_0=1$ and in the figure shown corresponds to $t_{today} \simeq
1.07$. The input values for the densities are $\rho_b=0.238,
\rho_{e}=0.0$ and $w_e$ is 0.5. The dilaton value today is taken
$\phi=0.0$ .  Right panel: The values of $\Omega_i \equiv \rho_i/\rho_c$
for the various species as functions of $t_{Einstein}$.  }
\label{fig12}  
\end{figure}

\begin{figure}[ht] 
\begin{center}
\includegraphics[width=4cm]{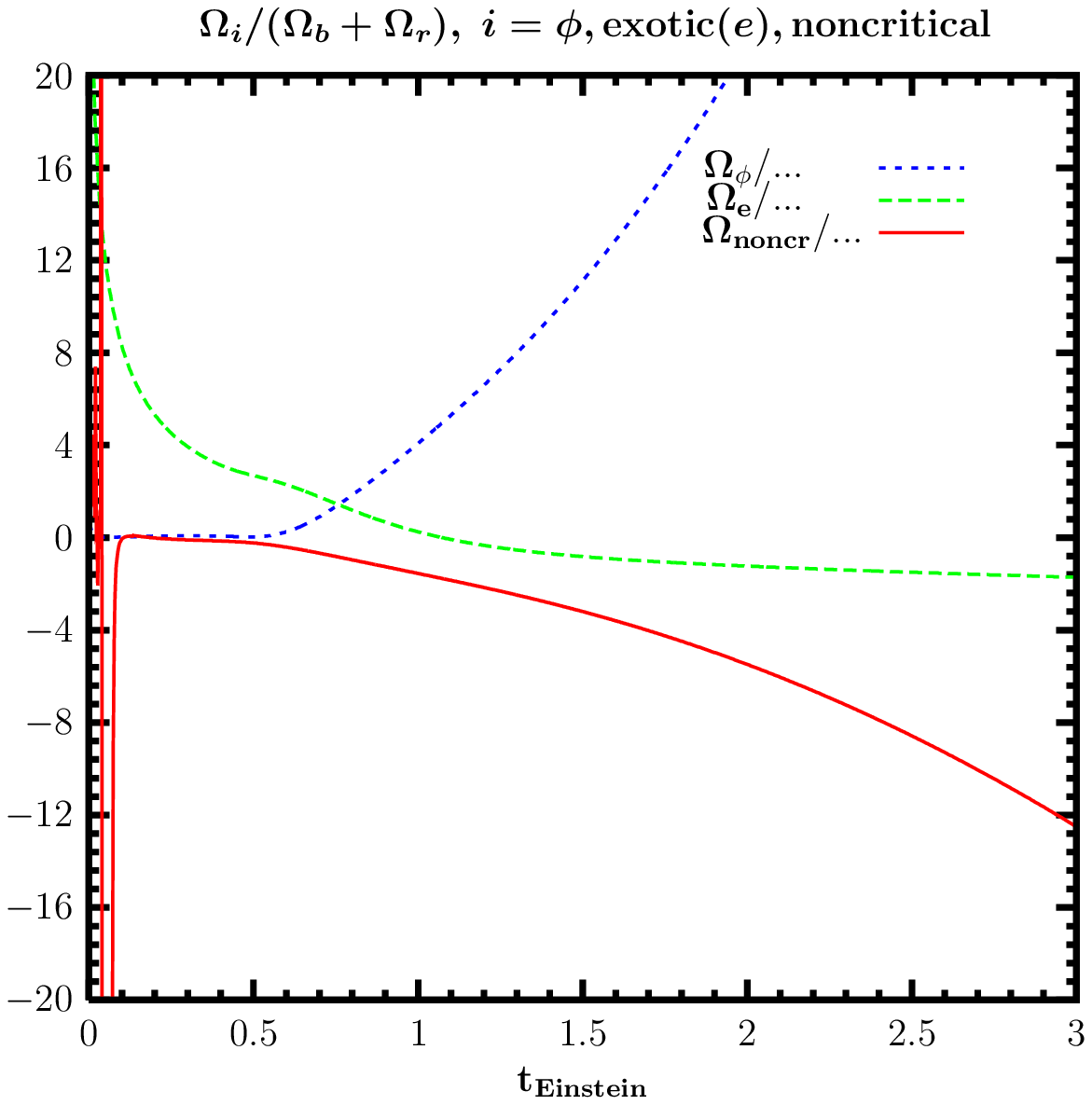}
\includegraphics[width=4cm]{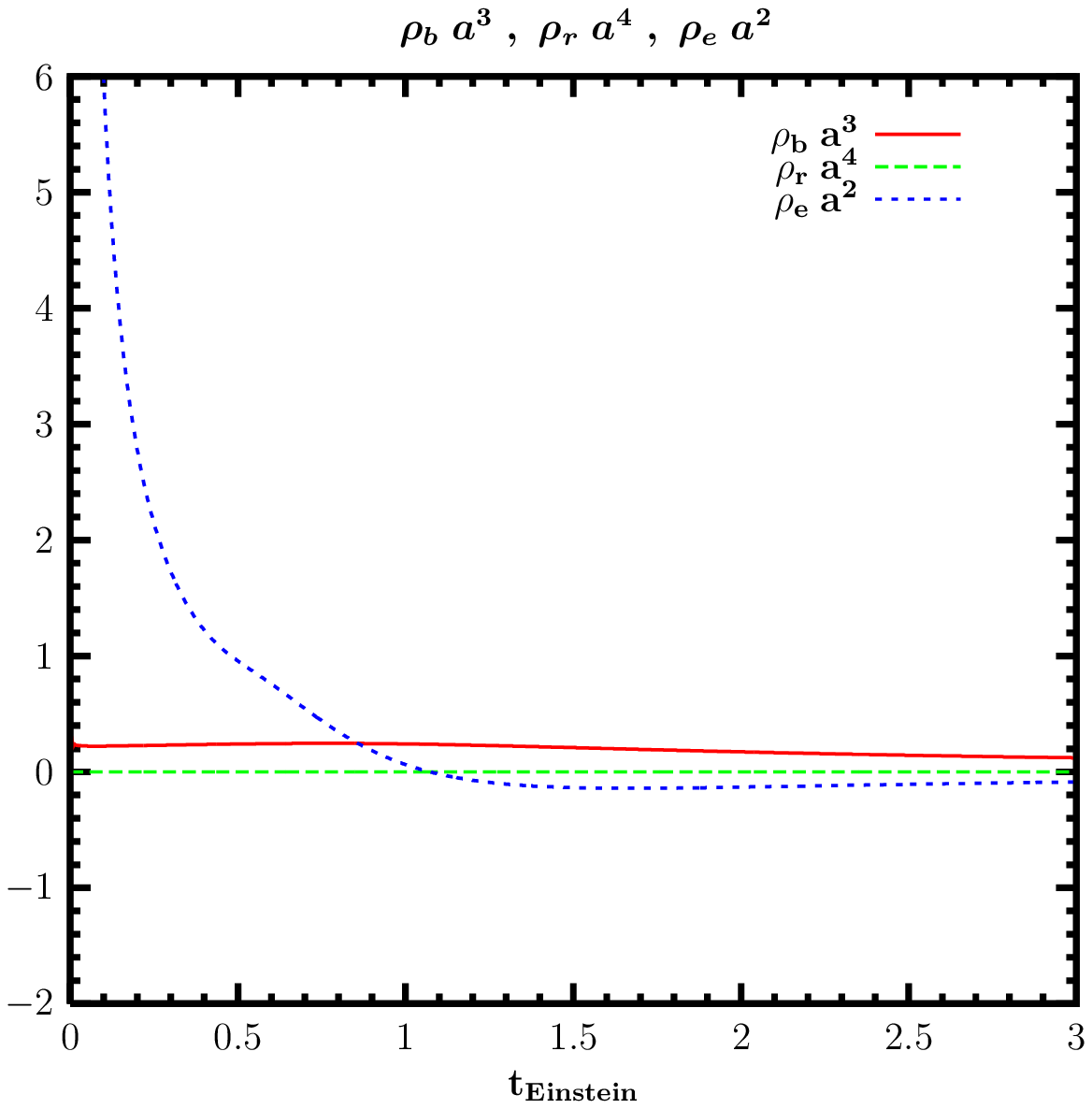}
\end{center}
\caption[]{ Left panel: Ratios of $\Omega$'s for the dilaton ($\phi$), exotic
matter ($e$) and the non-critical terms ("noncrit'') to the sum of "dust'' and radiation 
$\Omega_b + \Omega_r$ densities. \\
Right panel: The quantities $\rho_b \;a^3$, for "dust", $ \;\rho_r
\;a^4$ and $\rho_{e} \;a^2$ as functions of $t_{Einstein}$.  }
\label{fig34}  
\end{figure}

\begin{figure}[ht] 
\begin{center}
\includegraphics[width=4cm]{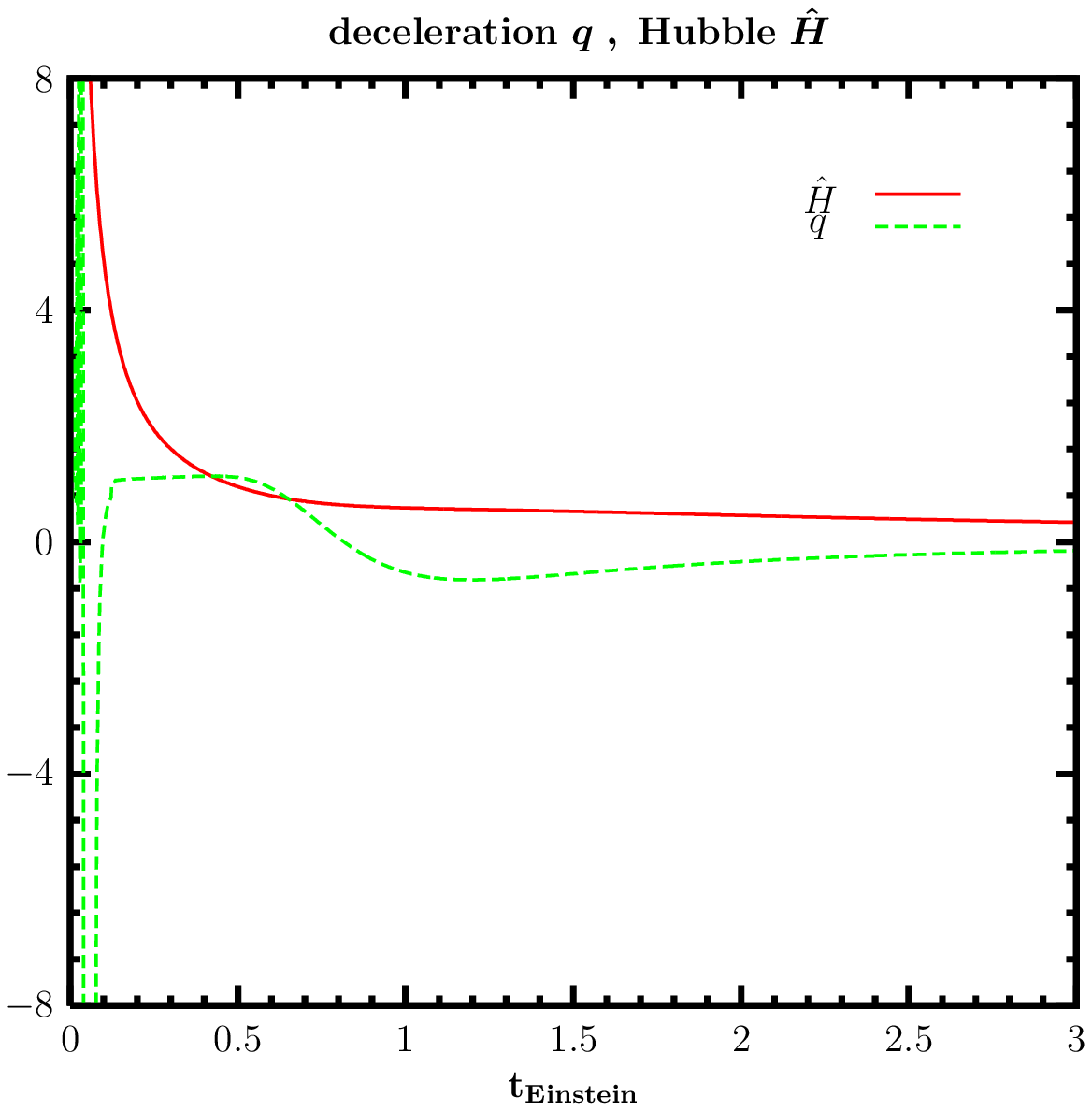}
\includegraphics[width=4cm]{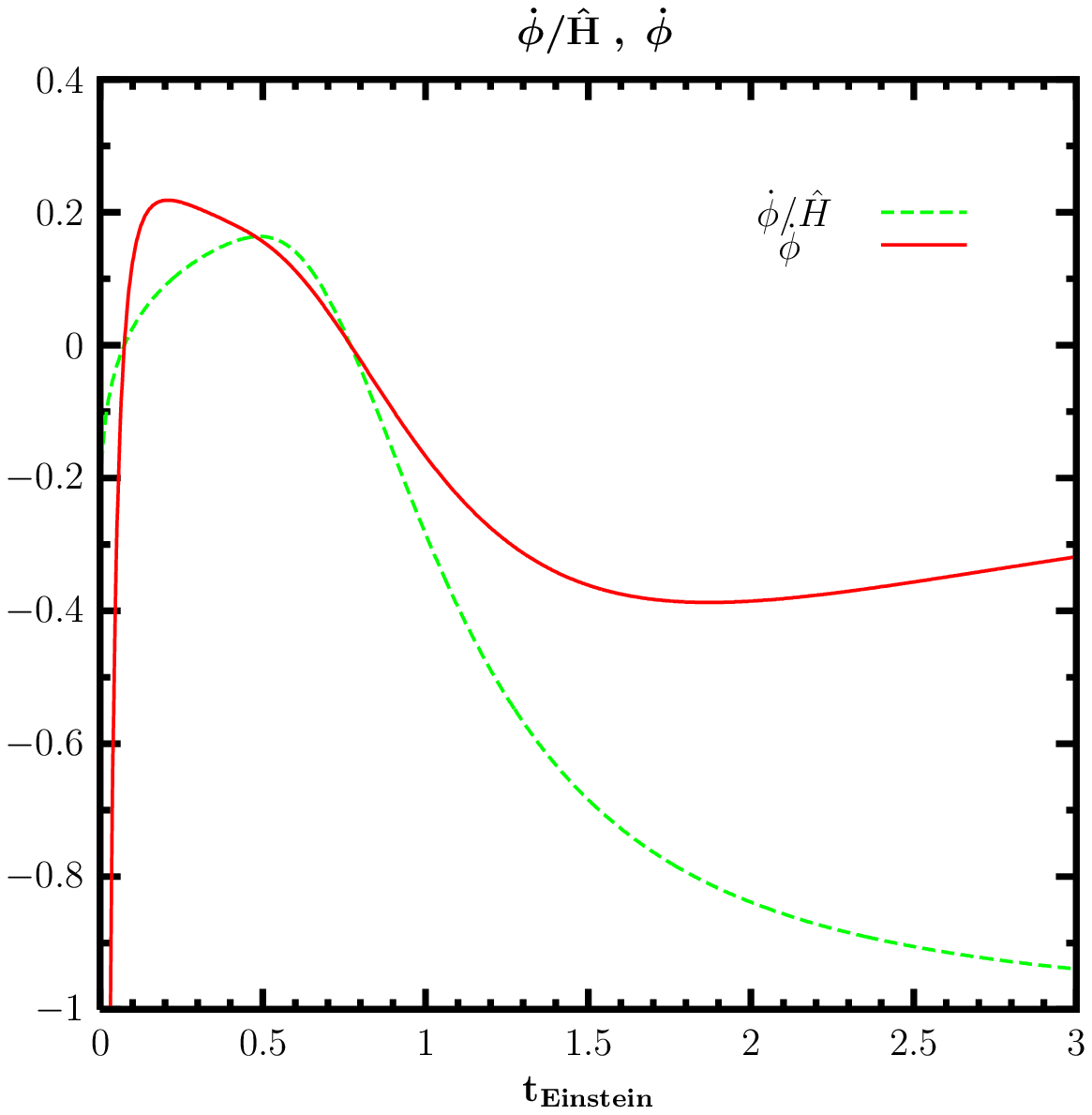}
\end{center}
\caption[]{ Left panel: The deceleration $q$ and the dimensionless
Hubble expansion rate $\hat{H}\equiv \frac{H}{\sqrt{3} H_0}$ as
functions of $t_{Einstein}$.  Right panel : The derivative of the
dilaton and its ratio to the dimensionless expansion rate.  }
\label{fig5}  
\end{figure}

\begin{figure}[ht] 
\begin{center}
\includegraphics[width=4cm,height=4.16cm]{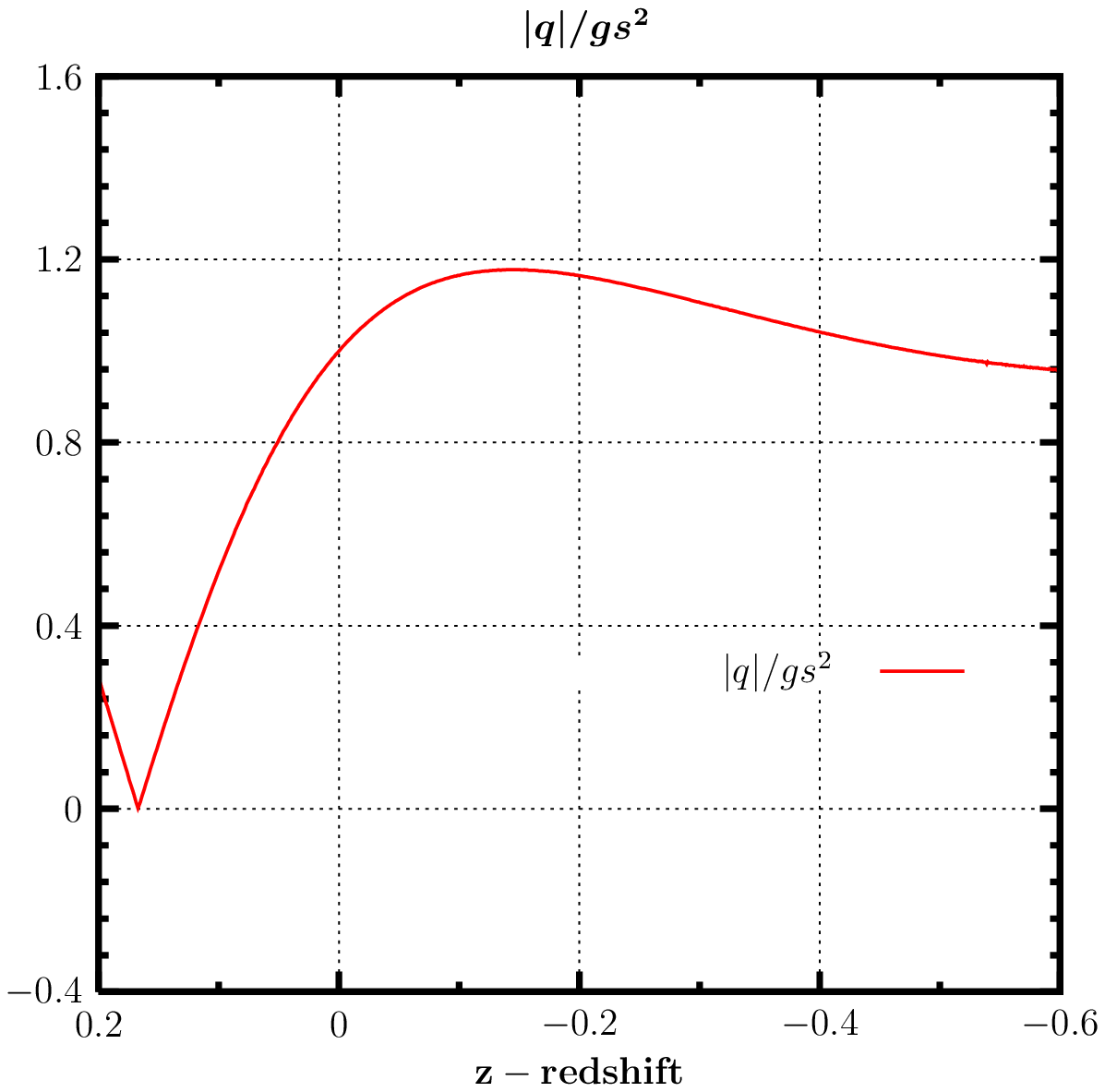}
\includegraphics[width=4cm]{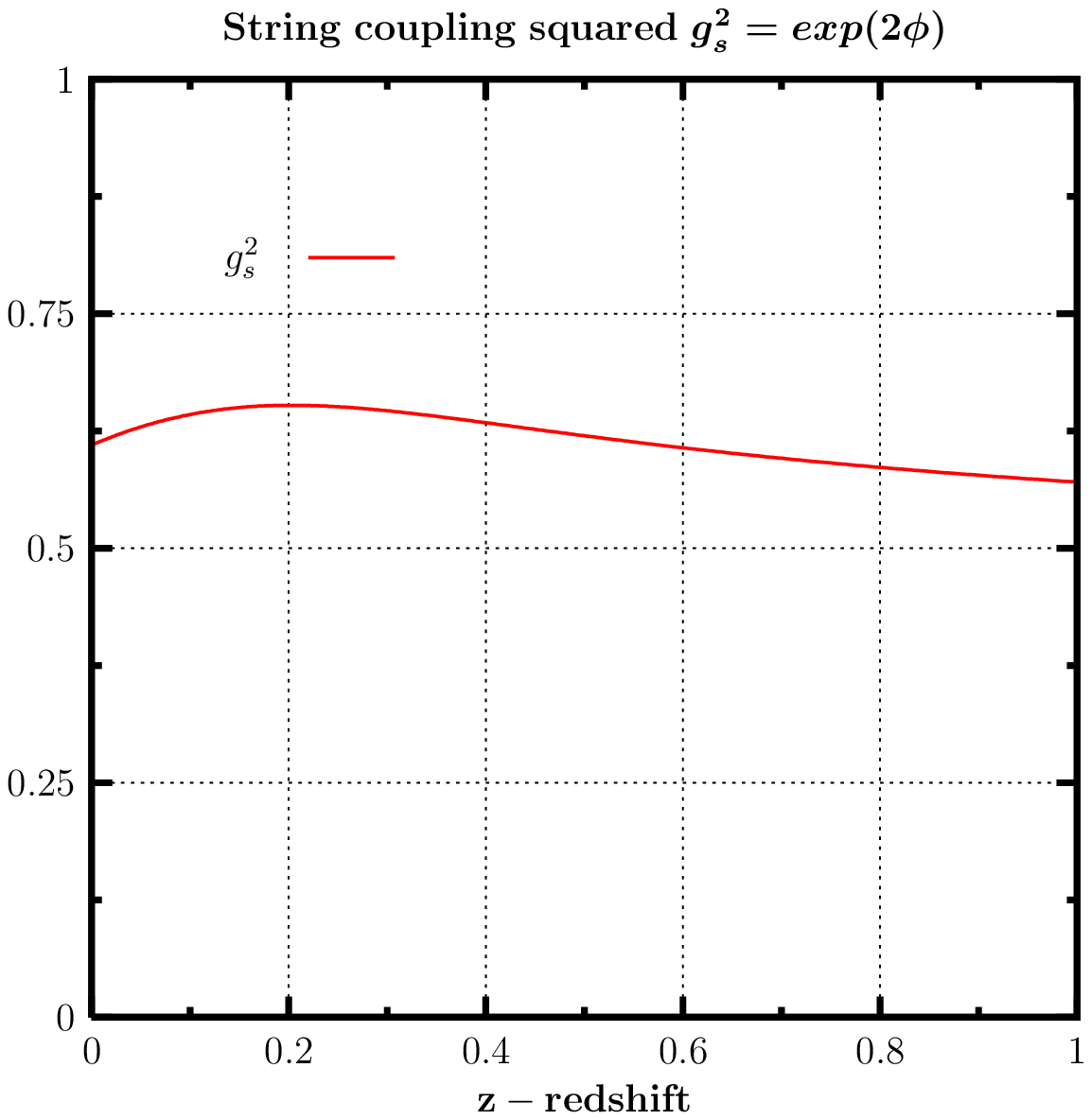}
\end{center}
\caption[]{ 
Left panel: 
The ratio $|q|/g_s^2$ as function of the redshift for $z$ ranging from $z=0.2$ 
to future values $z=-0.6$, for the inputs discussed in the main text. 
The rapid change near $z\approx 0.16$ signals the 
passage from deceleration to the acceleration period. 
Right panel: The values of the string coupling constant plotted 
versus the redshift in the range $z=0.0-1.0$. 
}
\label{qasz}  
\end{figure}

A brief summary of 
the results of our analysis for a model-case Q-cosmology, 
are presented 
in figs. \ref{fig12}, \ref{fig34}, \ref{fig5} and \ref{qasz}.
The model is discussed
in some detail in ref.~\cite{emn04}.  
Notice the late-era presence of exotic $a^{-2}$-scaling 
of matter species, attributed to dark matter, denoted by $\rho_e$ 
in the figures. 

The reader is invited to compare these results with the ones
of critical-string dilaton cosmologies in pre-Big-bang scenaria 
presented above (c.f. figs. 4 and 5), in particular 
with respect to the effects of the non-critical, off-shell 
terms ``${\cal G}$'',
which appear significant at the current era~\cite{emn04}. 

%
%

An important result of the analysis of \cite{emn04} 
is the fact that  
the conventional Boltzmann 
equation, controlling the evolution of species densities,  
needs to be modified in Q-cosmology, in 
order to incorporate consistently 
the effects of the dilaton {\it dissipative} pressure $\sim \dot{\phi}$
and the non-critical (relaxation) terms, $`` \cal{G}"$:
\begin{eqnarray}
\frac{dn}{dt}\;=\; - 3 \;H\; n\;-\;<\sigma v>\;( n^2-n^2_{eq}) 
{{\;+\;\dot{\phi} \; n + `` {\cal{G}} / {m_X} "}} \; \; .
\label{boltzman}
\end{eqnarray}

The respective relic density of the species $X$, with mass $m_X$, 
is then obtained from $\; \Omega_X
\; h_0^2= n\; m_X h^2_0$, after solving this modified equation. 
This may have important phenomenological consequences, 
in particular when obrtaining constraints on supersymmetric 
particle-physics models from astrophysical data.

%

We shall not discuss these issues further here, due to lack of space. 
For more details we refer the interested reader to the 
literature~\cite{emn,emn04}.

\section{Conclusions}

In this work we have reviewed various 
issues related to the consistent incorporation of Dark Energy in string theory.
We have discussed only traditional string theory and did not cover 
the modern extension, including membranes. This topic
has been covered by other lecturers in the School. 

One of the most important issues concerns de 
Sitter space, and in general space-times with horizons in string
theory. We have studied general properties, 
including holographic scenaria, which may be the key to an 
inclusion of such space times in the set of consistent 
(possibly non perturbative) ground states of strings.

We have also 
seen that perturbative strings are incompatible 
with space times with horizons, mainly due to the lack of a scattering 
matrix. However, non-critical strings may evade this constraint,
and we have discussed briefly how accelerating universes 
can be incorporated in non critical (Liouville) strings. 
The use of Liouville strings to describe the evolution of our Universe is
natural, since non-critical strings are associated with non-equilibrium
situations which undoubtedly occurred in the early Universe.  

The dilaton played an important r\^ole in string cosmology, and 
we have seen how it can act as a quintessence field, responsible 
for the current-era acceleration of the Universe.

There are many phenomenological tests of this class of cosmologies that
can be performed, which the generic analysis presented here is not
sufficient to encapsulate. Tensor perturbations in the cosmic microwave
background radiation is one of them. The emission of gravitational degrees
of freedom from the hot brane to the cold bulk, during the inflationary
and post-inflationary phases in models involving brane
worlds is something to be investigated in detail. 
A detailed knowledge of the dependence of the equation of state on the
redshift is something that needs to be looked at in the context of
specific models. 
Moreover, issues regarding the delicate balance of the expansion of the
Universe and nucleosynthesis, which requires a very low vacuum energy,
must be resolved in specific, phenomenologically semi-realistic models,
after proper compactification to three spatial dimensions, in order that
the conjectured cosmological evolution has a chance of success.

Finally, the compactification issue \emph{per se} is a most important part
of a realistic stringy cosmology. In our discussion above, we have
assumed that a 
consistent 
compactification takes place, leading to effective 
four-dimensional string-inspired equations of motion.
In realistic scenaria, however, details of how the extra dimensions
are compactified play a key r\^ole in issues like 
supersymmetry breaking.

In this review I did not discuss higher-curvature modifications 
of the low-energy Einstein action, which characterise 
all string-inspired models, including brane worlds scenaria.
Such terms may play an important r\^ole in Early Universe 
cosmology. For instance, they may imply initial singularity-free 
string cosmologies~\cite{tamvarizos},
or non-trivial black hole solutions with (secondary) dilaton hair~\cite{kanti},
which can play a r\^ole in the Early universe sphaleron transitions.
So, before closing the lecture, I will devote a few words on their form. 

In ordinary string theory, which is the subject of the present 
lecture, such higher-order terms possess ambiguous coefficients
in the effective action. This is a result of local field redefinitions,
which leave the (low-energy) string scattering amplitudes  
invariant, and hence cannot be determined by low energy considerations.
In ordinary string theory~\cite{strings}, with no space-time boundaries in
(the low-energy) target space time,  
such ambiguities imply that the so-called 
ghost-free Gauss-Bonnet combination 
$\frac{1}{g_s^2}\left(R_{\mu\nu\rho\sigma}^2 - 4 R_{\mu\nu}^2 + R^2\right)$,
with $g_s = e^\Phi$ the string coupling and $\Phi$ the dilaton field,   
can always 
be achieved  
for the quadratic curvature terms in the string-inspired low-energy 
effective action, which constitutes  
the first non trivial order corrections to Einstein 
term 
in bosonic and heterotic string effective actions. 

However, in the case of brane worlds, with closed strings propagating 
in the bulk, things are not so simple. As discussed in \cite{mp},
field redefinition ambiguities for the bulk low-energy 
graviton and dilaton fields, that would otherwise leave 
bulk string scattering amplitudes invariant, induce  
brane (boundary) curvature and cosmological constant 
terms, with the unavoidable result of 
ambiguities in the terms defining the Einstein and cosmological constant 
terms on the brane. This results in (perturbative in $\alpha '$) 
ambiguities  in the cross-over scale of four-dimensional brane gravity,
as well as the brane vacuum energy. It is not clear to me, however, 
whether these
ambiguities are actually present in low-energy brane world scenaria. 
I believe that these bulk-string 
ambiguities can be eliminated once the brane effective theory 
is propertly defined, 
given that closed and open strings also propagate on the brane world 
hypersurfaces, and thus are characterised by their own scattering amplitudes. 
Matching these two sets of scattering amplitudes properly, for instance 
by looking at the conformal theory describing the splitting of 
a closed-string bulk state, crossing a brane boundary, into two open 
string excitations on the brane, may lead to unambiguous 
brane cross-over and cosmological constant scales, expressed in terms of 
the bulk string scale and coupling.  
These are issues that I believe deserve further investigation, since they 
affect early Universe cosmologies, where such higher-curvature terms 
are important. I will not, however, discuss them further here. 

I would like to close this lecture 
with one more remark on the non-equilibrium
Liouville approach to cosmology advocated in \cite{emn,emn04}, 
and discussed last in this article. This 
approach is based
exclusively on the treatment of target time as an irreversible dynamical
renormalization-group scale on the world sheet of the Liouville string
(the zero mode of the Liouville field itself). This irreversibility is
associated with fundamental properties of the world-sheet renormalization
group, which lead in turn to the loss of information carried by
two-dimensional degrees of freedom with world-sheet momenta beyond the
ultraviolet cutoff~\cite{zam} of the world-sheet theory. This fundamental
microscopic time irreversibility may have other important consequences,
associated with fundamental violations of CPT
invariance~\cite{mavrodecoh} in both the early Universe
and the laboratory, providing other tests of these ideas.

\noindent 
{\bf Acknowledgements} \\ 
It is my pleasure to 
thank the organisers, and especially E. Papantonopoulos, 
for the invitation to lecture in this 
very interesting school and workshop. 
This work is partially supported by funds made available 
by the European Social Fund (75\%) and National (Greek)
Resources (25\%) - (EPEAEK II) - PYTHAGORAS. 


\end{document}